\documentclass[ 
                reprint,
               showpacs, 
         amsmath,amssymb,aps,prd,
              nofootinbib,eqsecnum, amsart,
                superscriptaddress,
                longbibliography,
        floatfix ,
               a4paper 
               ]{revtex4-1}

\usepackage{hyperref}
\usepackage{graphicx}
\usepackage{xcolor}
\usepackage{bm}
\usepackage{afterpage}

\usepackage{natbib}
\usepackage[english]{babel}
\usepackage{dcolumn}
\DeclareMathVersion{nxbold} 
\SetSymbolFont{operators}{nxbold}{OT1}{cmr} {b}{n}
\SetSymbolFont{letters}  {nxbold}{OML}{cmm} {b}{it}
\SetSymbolFont{symbols}  {nxbold}{OMS}{cmsy}{b}{n}

\usepackage{todonotes}

\input{_jnls.cls} 

\usepackage{siunitx}
\usepackage{adjustbox}
\usepackage{enumitem}
\usepackage{afterpage}
\usepackage{currfile}

\DeclareMathAlphabet{\mathpzc}{OT1}{pzc}{m}{it}

\newcommand{\infrac}[2]{#1/#2}
\newcommand{\arm}{a_\text{eq}}
\newcommand{\Hrm}{\mathcal{H}_\text{eq}}

\renewcommand{\vec}[1]{\mathbf{#1}}
\newcommand{\Meszaros}{M{\'e}sz{\'a}ros }
\newcommand{\eref}[1]{Eq.~(\ref{#1})}
\newcommand{\steref}[1]{Equation~(\ref{#1})}
\renewcommand{\eqref}[1]{(\ref{#1})}
\newcommand{\vecf}{\vec{f}}
\newcommand{\vecp}{\vec{p}}

\newcommand{\matA}{\vec{A}}
\newcommand{\matB}[1][]{\vec{B}_{#1}}
\newcommand{\matC}{\vec{C}}
\newcommand*\nc[1]{\tikz[baseline=(char.base)]{
		\node[shape=circle,draw,inner sep=2pt] (char) {\small #1};}}
\newcommand*\tnc[1]{\tikz[baseline=(char.base)]{
    \node[shape=circle,draw,inner sep=1pt] (char) {\small #1};}}
\newcommand{\balg}{\begin{enumerate}[label=\protect\nc{\arabic*}]}
\newcommand{\ealg}{\vspace{6 pt}\end{enumerate}}

\newcommand{\smax}{s_\text{max}}

\newcommand{\dr}{\delta_\text{r}}
\newcommand{\dm}{\delta_\text{m}}

\newcommand{\tauc}{\tau_\text{c}}
\newcommand{\rhor}{\rho_\text{r}}
\newcommand{\rhom}{\rho_\text{m}}
\newcommand{\kunit}{\infrac{\Hrm}{\sqrt{2}}}
\newcommand{\tunit}{\infrac{\sqrt{2}}{\Hrm}}
\newcommand{\Prr}{P_\text{r}}
\newcommand{\Vr}{V_\text{r}}
\newcommand{\Vm}{V_\text{m}}

\newcommand{\rcell}[1]{\multicolumn{1}{r}{#1}}
\newcommand{\drs}{\delta_{\text{r}(s)}}
\newcommand{\drsnum}{\delta_{\text{r}(s)}^\text{num}}
\newcommand{\smin}{S}

\newcommand{\rfcite}[1]{Ref.~\cite{#1}}

\DeclareMathOperator{\re}{Re}

\def\H{{\cal{H}}}
\def\bea{\begin{eqnarray}}
\def\eea{\end{eqnarray}}
\def\bi{\begin{itemize}}
\def\ei{\end{itemize}}
\def\be#1\ee{\begin{equation}#1\end{equation}}
\def\ba#1\ea{\begingroup
\addtolength{\jot}{9pt}\begin{align}#1\end{align}\endgroup}
\def\bas#1\eas{\begingroup
\addtolength{\jot}{9pt}\begin{align}\begin{split}#1\end{split}\end{align}\endgroup}
\def\bfa#1\efa{\begingroup
\addtolength{\jot}{9pt}\begin{flalign}#1\end{flalign}\endgroup}
\def\bml#1\eml{\bgroup
\addtolength{\jot}{9pt}\begin{multline}#1\end{multline}\egroup}

\def\nt#1\ent{#1 }

\newcommand{\normalised}[1]{#1}
\newcommand{\kn}{\normalised{k}}
\newcommand{\tn}{\normalised{\tau}}
\newcommand{\diml}[1]{\breve{#1}}
\newcommand{\kd}{\diml{k}}
\newcommand{\td}{\diml{\tau}}
\newcommand{\trm}{\tn_\text{eq}}

\begin{document}

\makeatletter
\newcolumntype{Z}[3]{>{\mathversion{nxbold}\DC@{#1}{#2}{#3} }c<{\DC@end} } 
\makeatother
\newcommand{\bcell}[1]{\multicolumn{1}{Z{.}{.}{-1} }{#1}}

 
\title{Double power series method for approximating cosmological perturbations}

\author{Andrew~J.~Wren}
\email[]{andrew.wren@ntlworld.com}
\affiliation{Astronomy Unit, School of Physics and Astronomy, Queen Mary University of London, Mile End Road, London, E1 4NS, United Kingdom}
\author{Karim~A.~Malik} \email[]{k.malik@qmul.ac.uk}
\affiliation{Astronomy Unit, School of Physics and Astronomy, Queen Mary University of London, Mile End Road, London, E1 4NS, United Kingdom}

\date{\today}

\begin{abstract}
We introduce a double power series method for finding approximate
analytical solutions for systems of differential equations commonly
found in cosmological perturbation theory.  The method was set out, in
a non--cosmological context, by Feshchenko, Shkil' and Nikolenko (FSN)
in 1966, and is applicable to cases where perturbations are
on sub--horizon scales.
The FSN method is essentially an extension of the well known
Wentzel-–Kramers-–Brillouin (WKB) method for finding approximate analytical
solutions for ordinary differential equations. The FSN method  we use is applicable well beyond perturbation theory to solve systems of ordinary differential equations, linear in the derivatives, that also depend on a small
parameter, which here we take to be related to the inverse wave--number.

We use the FSN method to find new approximate oscillating
solutions in linear order cosmological perturbation theory for a flat
radiation--matter universe.  Together with this model's well known growing and
decaying \Meszaros solutions, these oscillating modes provide a
complete set of sub--horizon approximations for the metric potential, radiation
and matter perturbations. Comparison with numerical solutions of the perturbation equations shows that our approximations can be made accurate to within a typical error of $1\%,$ or better. We also set out a heuristic method for error estimation.
A \emph{Mathematica} notebook which implements the double power series method is made available online.
\end{abstract}

\pacs{ 
98.80.Jk, 
02.30.Mv, 
 04.25.Nx, 
95.30.Sf 
\hfill } \preprint{arXiv:1603.07577 [gr-qc]}

\maketitle

\section{Introduction}
\label{sec:intro}

A hundred years after Einstein published his theory of General
Relativity, there are still only a fairly limited number of exact
solutions to the field equations known. In order to solve the
governing equations, the community is relying on approximate solution
schemes, like perturbation theory, which are then solved numerically.
However, there is still room for approximate analytical solutions of
the perturbed equations. These can test numerical solutions,
increase efficiency if numerical solutions are computationally costly
or provide input for non--linear perturbation
equations. Approximate analytical solutions also provide insight which
can be difficult to gain from numerical results. For example,
analytical approximations can highlight that certain perturbation
equations have both oscillating and non--oscillating short wavelength
(sub--horizon) modes.

Arguably, the most familiar approximate analytical result is that a sub--horizon matter perturbation in an  Einstein--de Sitter universe has a growing mode proportional to the scale factor, $a$, and a decaying mode proportional to $a^{-3/2}$ \cite{Lifshitz1946,1970ApJ...162..815P}. In the Einstein-de Sitter universe, it is also possible to write down an \emph{exact} analytical expression for these perturbations, which does not depend on the sub--horizon approximation.  Approximate analytical results are also useful in the more complicated setting of a radiation--matter universe~--- as reviewed in \rfcite{2006AIPC..843..111P}, for example.  For more information on sub--horizon non--oscillating modes in such a universe see, for example, Refs.~\cite{1974A&A....37..225M,1975A&A....41..143G,2002ApJ...581..810W}. For the sub--horizon oscillating modes, see, for example, Refs.~\cite{1970ApJ...162..815P,2002ApJ...581..810W}. Analytical approximations have also been used in the context of other models of the Universe~--- for example, Refs.~\cite{1992ARA&A..30..499C,chernin2003growth} derive results for non--oscillating modes in a $\Lambda$CDM universe, while Refs.~\cite{Martin:2002vn, Casadio:2004ru} consider WKB--type approximation of perturbations during cosmic inflation. \\

In this paper, we concentrate on oscillating modes in a flat radiation--matter universe, focusing on improving the accuracy of approximate analytical solutions to the perturbed governing equations.  The matter component is pressureless dust, and our perturbation equations treat both radiation and matter as perfect fluids. As we shall see, the double power series method we use provides a considerable improvement in accuracy of modelling the gravitational interactions over leading order approximations. We believe our method is new to the cosmological literature.

More specifically, we derive approximate analytical solutions to linear scalar
cosmological perturbation equations.  Our analytical solutions only
involve simple functions such as polynomials and exponentials. They are valid
in the sub--horizon approximation~--- that is for perturbations with
co--moving wave--number significantly larger than the conformal Hubble
parameter.

Section~\ref{sec:introducing-the-double-power-series-method}
introduces the double power series method. It was first set out in~\rfcite{FSN}, in a non--cosmological
context.  The double power series method is based on constructing an
approximate solution as the product of a first power series with the
\emph{exponent} of a second power series~--- this is set out below in
\eref{eq:xexpn}. The exponentiated power series enables the method to
deal effectively with oscillating solutions. Before employing
the method in a cosmological context, we illustrates its use for a
simple Bessel equation.  Appendix~\ref{sec:Bessel-calculation} goes through the details of the underlying calculation.

In Section~\ref{sec:perturb}, we look at a model where the energy density of the universe is due only to radiation and pressureless matter.  This model reflects the make up of the actual Universe from the time of decoupling onwards, until the much later epoch when dark energy becomes significant. Using the formalism of~\rfcite{2009PhR...475....1M},  we set out the equations which govern radiation and pressureless matter perturbations in the longitudinal gauge. We choose to focus on a pair of second order differential equations for the metric potential and radiation perturbations.  The use of this pair of equations neatly captures that there are four independent solutions, or modes, for the perturbations.  We also note in Appendix~\ref{sec:double-power-series-for-perturbations} that, in a precise sense, the oscillating modes are driven by radiation pressure, while the growing and decaying modes are associated with presence of pressureless matter.

In Section~\ref{sec:leading-order}, we review the leading order approximations to both oscillating and non-oscillating modes, deriving them using the second order differential equations of Section~\ref{sec:perturb}. For the oscillating modes, this is a  leading order  Wentzel-–Kramers-–Brillouin (WKB) approximation, along lines found in \rfcite{1970ApJ...162..815P}~--- our use of the double power series method extends these approximations to higher orders.  The leading order growing and decaying mode approximations are the \Meszaros solutions~\cite{1974A&A....37..225M,1975A&A....41..143G}.  When we describe an approximation as $n$th order, this should not be confused with the order of the perturbation theory as described in, for example, \rfcite{2009PhR...475....1M}. It refers instead to the order of approximation in solving the differential equations. 

Our main focus in this paper is on the oscillating modes~--- to the best of our knowledge, our higher order results are not found in the literature. Section~\ref{sec:prep} puts the linear perturbation equations in the most helpful form for applying the double power series method, expressing the key pair of equations from Section~\ref{sec:perturb} explicitly in terms of conformal time and then presenting them as a single $2\times 2$ matrix equation. Next, in Section~\ref{sec:an-approximate-solution-for-first-order-scalar-fast-modes-in-the-flat-radiation--matter-model}, we use the double power series method to derive our approximate oscillating solutions. We use the method directly to approximate the radiation perturbation $\dr$ and the scalar potential perturbation $\psi.$  We then derive the corresponding approximation for $\dm.$ The approximate solution to third order is calculated in detail in Appendix~\ref{sec:double-power-series-for-perturbations} and set out in \eref{eq:approx}. Appendix~\ref{sec:app-seventh} presents the solution to seventh order. For perturbations which are sufficiently sub--horizon, we can get approximations accurate to a typical error of $1\%$ or better compared with the exact numerical solution of the perturbation equations. The order of double power series needed to get this level of accuracy depends mainly on the wave--number of the perturbation and the range of times which are of interest.

Section~\ref{sec:error-estimates} sets out a heuristic method for estimating the error without calculating the numerical solution.  This heuristic also provides a method of identify the optimal order of approximation for a given wave--number and conformal time.
 
Section~\ref{sec:conclusion} concludes the paper. It summarises the results and provides a brief discussion.  We note that, while the leading order WKB approximation suggests the oscillating modes have a constant period, in fact this period decreases with time.  The conclusion also sets out some ideas on further use of the double power series method in cosmology.

A \emph{Mathematica} notebook which executes the double power series method automatically is available online at \rfcite{Github}.  It enables all the double power series in this paper to be calculated in at most a few seconds on a standard PC.  The template can be easily adapted for use with similar systems of second order differential equations.

\section{The double power series method}
\label{sec:introducing-the-double-power-series-method}

In this section we introduce the double power series method. We
explain briefly its motivation and give an example of its use for a
simple ordinary differential equation of Bessel type.

Systems of differential equations in cosmological perturbation theory
tend to contain coefficients involving quantities such as the
conformal Hubble parameter $\H$ and the co--moving wave--number. (For
example, see Eqs.~\eqref{eq:rgeq1c} and~\eqref{eq:rgeq2c} below.)
These quantities provide two different timescales for the evolution of
perturbations.   We will write the wave--number as $\kd,$ with the
accent (a ``breve'') indicating that this is a dimensionful
quantity~--- soon we will normalise this to a dimensionless number.
 We focus on the sub--horizon case, $\kd\gg\H.$  $\kd$ and $\H$ then
define timescales $(\kd c)^{-1}$ and $(\H c)^{-1},$ where $c$ is the
speed of light, which, as is conventional, we will set to be equal to
$1.$ The sub--horizon condition that implies the timescale associated
with $\kd$ is much shorter than the timescale associated with $\H.$

There are various methods to deal with multiple timescales. The usual approach is reviewed in the standard reference work~\cite{1978amms.book.....B}. An alternative method which we found useful for cosmological perturbation theory is set out in~Ref.~\cite[Chapter~2]{FSN}.  To the best of our knowledge, this paper represents the first time \rfcite{FSN}'s method has been used in cosmology. \\

To explain the method, it is convenient to express the system of
equations as a single matrix equation,
\be 
\label{eq:matrxversion}
\matA\vecf''(\td)+\matC\vecf'(\td)+\matB\vecf(\td)=\vec{0}\,,
\ee
where we have used  $\td$ for the time variable since in this paper
we will work in conformal time, and because, as for the wave--number
we will subsequently choose a normalised dimensionless conformal time
(a ``dash'' indicates as usual differentiation with respect to the
independent variable).  The matrices $\matA,\matC$ and $\matB$
depend on $\H\sim\td^{-1}$ and $\kd.$ The approximation relies on the
equations involving two very different rates of change, a slow rate of
change, of order $\H\sim\td^{-1},$ and a fast rate of change $\kd.$ In
terms of  the  cosmological perturbation theory equations,
these rates are, respectively, associated with the relatively slow
expansion of the Universe,  associated with $\H,$  and the
relatively fast  fluctuation  of sub--horizon waves, 
associated with $\kd.$  The fast rate of change dominates
oscillating solutions to the equations, and we will find that an
approximate solution can be  derived order by order  in
descending powers of  the wave--number.

 In our approximation, we will choose some reference conformal time $\tauc$ (see especially \eref{eq:taucrit} below) and define dimensionless parameters $\tn=\infrac{\td}{\tauc}$ and $\kn=\kd\tauc$ with $\kn\gg\tn^{-1},$  and hence $\kn^{-1}$ being a small parameter in the sense that $\kn^{-1}\ll\tn.$    The double power series method of Ref.~\cite[Chapter 2]{FSN} finds a series approximation for $\vecf$,
\be \label{eq:xexpn}
\vecf(\tn)=\left(\sum_{j=0}^\infty \kn^{-j}\, \mathbf{p}_j(\tn)\right) \exp\left[\sum_{n=0}^\infty \int \kn^{-n+1}\,\omega_n(\tn)\,d\tn\right]\,,
\ee
where $\mathbf{p}_j(\tn)$ and $\omega_n(\tn)$ are independent of $\kn,$  and are assumed to have slow rates of change like $\tn^{-1}\ll\kn.$ The notation of \rfcite{FSN} is superficially different to ours, and a comparison between them is provided in  Table~\ref{table:FSNdictionary}.

\begin{table*}
\bgroup
\begin{ruledtabular}
\def\arraystretch{1.5}
\addtolength{\jot}{9pt}
\caption[Our notation and \rfcite{FSN}'s compared]{The notation used in this paper compared with the notation introduced in Chapters~1 and 2 of \rfcite{FSN}. Since our differential equations are all homogeneous~--- in the sense that a multiple of a solution is also a solution~--- so we use the ``non-resonance'' case from \rfcite{FSN}.\label{table:FSNdictionary}}
\begin{tabular}{lccccccccccccccc}
{This paper}  & $\td$ and $\tn$  & $\kn$ &$\eta=\kn\tn$ & $f$ &$A$ &$B_{-2}$ & $B_0$ & $C$ & $\omega$&  $\vecf$ & $\matA$ &$\matB[-2]$ & $\matB[0]$ & $\matC$ & $\vecp$\\ 
\hline
\rfcite{FSN}& $\tau$ & $\epsilon^{-1}$ & $t$  & $x$ &$a_0$  & $b_0$   & $b_2$   & $c_0$ & $D+i\Omega$ &  $x$ &$A_0$  & $B_0$   & $B_2$   & $C_0$ & $\Pi$ \\ 
\end{tabular}
\end{ruledtabular}
\egroup
\end{table*}

The introduction of the exponential factor is the key element of the method. In the first term of \eref{eq:xexpn} only zero and negative powers of $\kn$ occur, while in the exponent, a power $\kn^1$ also occurs.  This is because, as noted before \eref{eq:xexpn}, $\kn^{-1}$ is a small parameter, in the sense that $\kn^{-1}\ll\tn.$ The $n=0$ term of $\sum_{n=0}^\infty \int \kn^{-n+1}\,\omega_n(\tn)\,d\tn$ is proportional to $\kn$ and gives the leading order of the frequency of $\vecf.$  The $n>0$ terms give smaller corrections to the frequency. Starting instead from $n=1,$ would give only low frequency oscillations, whereas in this paper we are principally interested in rapidly oscillating solutions.

 The decomposition set out in \eref{eq:xexpn} is widely applicable to differentiable functions which depend on some parameter $k.$ The success of the FSN method depends on the coefficient functions $\mathbf{p}_j(\tn)$ and $\omega_n(\tn)$ being slowly varying in $\tau,$ and the whole series being sufficiently well behaved, at least for low enough $j$ and $n.$ A successful  decomposition approximates the function $\vecf(\tn)$ sufficiently closely for a suitable range of conformal times $\tn.$  As we shall see in Section~\ref{sec:error-estimates}, for some values of $\tau$ the decomposition may be convergent, for others, uncontrollably non-convergent, and, for yet other values, have an intermediate property of being \emph{asymptotically convergent}. Asymptotic convergence is extensively discussed in, for example, \rfcite{1978amms.book.....B}.

 In passing, we note that a related, and more general, FSN method, discussed in Ref.~\cite[Chapter 3]{FSN}, can be used to solve systems of differential equations. This can include systems of coupled oscillators with differing frequencies.  However, for our present purposes, this is not necessary: our oscillators have only a single (complex) frequency, associated at time $\tau$ with $\omega(\tn)=\sum_{n=0}^\infty \int \kn^{-n+1}\,\omega_n(\tn)\,d\tn.$

The method now consists of substituting \eref{eq:xexpn} into~\eref{eq:matrxversion} and equating coefficients of  powers of $\kn$. This is most easily explained by illustrating the approach for a simpler equation of the same form.   \\

The matrices $\matA, \matC$ and $\matB$ of \eref{eq:matrxversion} are now replaced by scalar functions $A,C$ and $B$ with
\be  \label{eq:ABC-scalar}
A=1,\quad C=\frac{1}{\tn}\quad\text{and}\quad B=\kn^2+\frac{1}{\tn^2}\,.
\ee

The equation can be solved analytically~--- it is equivalent to one of the well known Bessel equations~(see, for example, 10.2.1 in the online reference manual \cite{NISTBessel}). To show this equivalence, we substitute \eref{eq:ABC-scalar} into \eref{eq:matrxversion} and multiply by $\tn^2$ to get
\be \label{eq:pre-bessel2}
\tn^2 f''(\tn)+\tn f'(\tn)+\left((\kn\tn)^2+1\right)f(\tn)=0.
\ee
Now, writing $\eta=\kn\tn,$ and noting that $\frac{d}{d\tn}=\kn\frac{d}{d\eta}$, we have
\be
\label{eq:bessel2}
\eta^2 \frac{d^2f}{d\eta^2}+\eta \frac{df}{d\eta}+\left(\eta^2+1\right) f=0.
\ee
This is the standard form of the Bessel equation of order $i=\sqrt{-1}$.  As set out in \rfcite{NISTBessel}, for example, the exact solution is therefore
\be
\label{eq:besselanalytic}
f(\tn)=C_J J_{i}(\kn\tn)+C_Y Y_{i}(\kn\tn),
\ee
where $C_J$ and $C_Y$ are arbitrary complex constants and $J_i$ and $Y_i$ are the Bessel functions with index $i$ of the first and second kind respectively.

We now re--write \eref{eq:pre-bessel2} as
\be
\label{eq:bessel-dls}
f''(\tn)+\frac{1}{\tn}f'(\tn)+\left(\kn^2+\frac{1}{\tn^2}\right)f(\tn)=0
\ee
and apply the double power series method using the one--dimensional version of \eref{eq:xexpn},
\bml \label{eq:besselexpn}
f(\tn)=
\left(\sum_{j=0}^\infty \kn^{-j}\, p_j(\tn)\right) \exp\left[{\sum_{n=0}^\infty \int \kn^{-n+1}\,\omega_n(\tn)\,d\tn}\right].
\eml
For convenience, write
\be 
\begin{aligned}\label{eq:POdefn}
	p&=\sum_{j=0}^\infty \kn^{-j}\, p_j(\tn),
	&\quad
	\omega&=\sum_{n=0}^\infty \kn^{-n+1}\,\omega_n(\tn)
\end{aligned}
\ee
and
\bml \label{eq:Edefn}
E=\exp\left[\int\omega(\tn)\,\mathrm{d}\tn\right]
\\
=\exp\left[\sum_{n=0}^\infty \int \kn^{-n+1}\,\omega_n(\tn)\,d\tn\right].
\eml
The terms $p, \omega$ and $E$ all depend on both $\tn$ and $\kn,$ and we will regard $\kn$ as a fixed parameter. We then have
\ba
f&=p E,\label{eq:f}\\
f'&=\left(p'+p\,\omega\right)E,\label{eq:fd}\\
\mathrm{and\quad\quad} f''&=\left(p''+2p'\,\omega+p\,\omega'+p\,\omega^2\right)E\,, 
\label{eq:fdd}
\ea
where the dashes denote differentiation with respect to $\tn.$  

Using the approximation \eref{eq:besselexpn} in our Bessel equation \eref{eq:bessel-dls}, and dividing through by $E,$ we get
\bml \label{eq:Besselapproxequation}
\left(p''+2p'\,\omega+p\,\omega'+p\,\omega^2\right)
\\
+\frac{1}{\tn}\left(p'+p\,\omega\right)+\left(\kn^2+\frac{1}{\tn^2}\right)p=0\,.
\eml

 Using $\kn\gg\tn^{-1},$ the leading order part of \eref{eq:Besselapproxequation} consists of the terms with the highest positive order $r$ of $\kn,$ which is $\kn^2,$ coming from the $p \, \omega^2$ term and the $\kn^2 p$ term. We therefore start by equating coefficients of $\kn^2.$ Feeding  these results back into \eref{eq:Besselapproxequation} gives us an equation with  the leading order part being of order $\kn.$   We equate coefficients of $\kn,$ and again feed the results of  that into \eref{eq:Besselapproxequation}.  We continue this process for coefficients of $1=\kn^0,$ then $\kn^{-1}$ and so on.  We stop the procedure after equating coefficients of $\kn^{2-\smax}$ for whatever value of $\smax\ge 0$ we choose: this value $\smax$ is the order of our approximation.

The overall algorithm is set out in more detail in
Figure~\ref{fig:Besselalgorithm}. We work through this algorithm in
Appendix~\ref{sec:Bessel-calculation}. We have also put a
\emph{Mathematica} notebook which executes this method on Github at
\rfcite{Github}, which can be used to calculate double power series
approximations easily.

\begin{figure} 
\centering
\includegraphics[width=\linewidth]{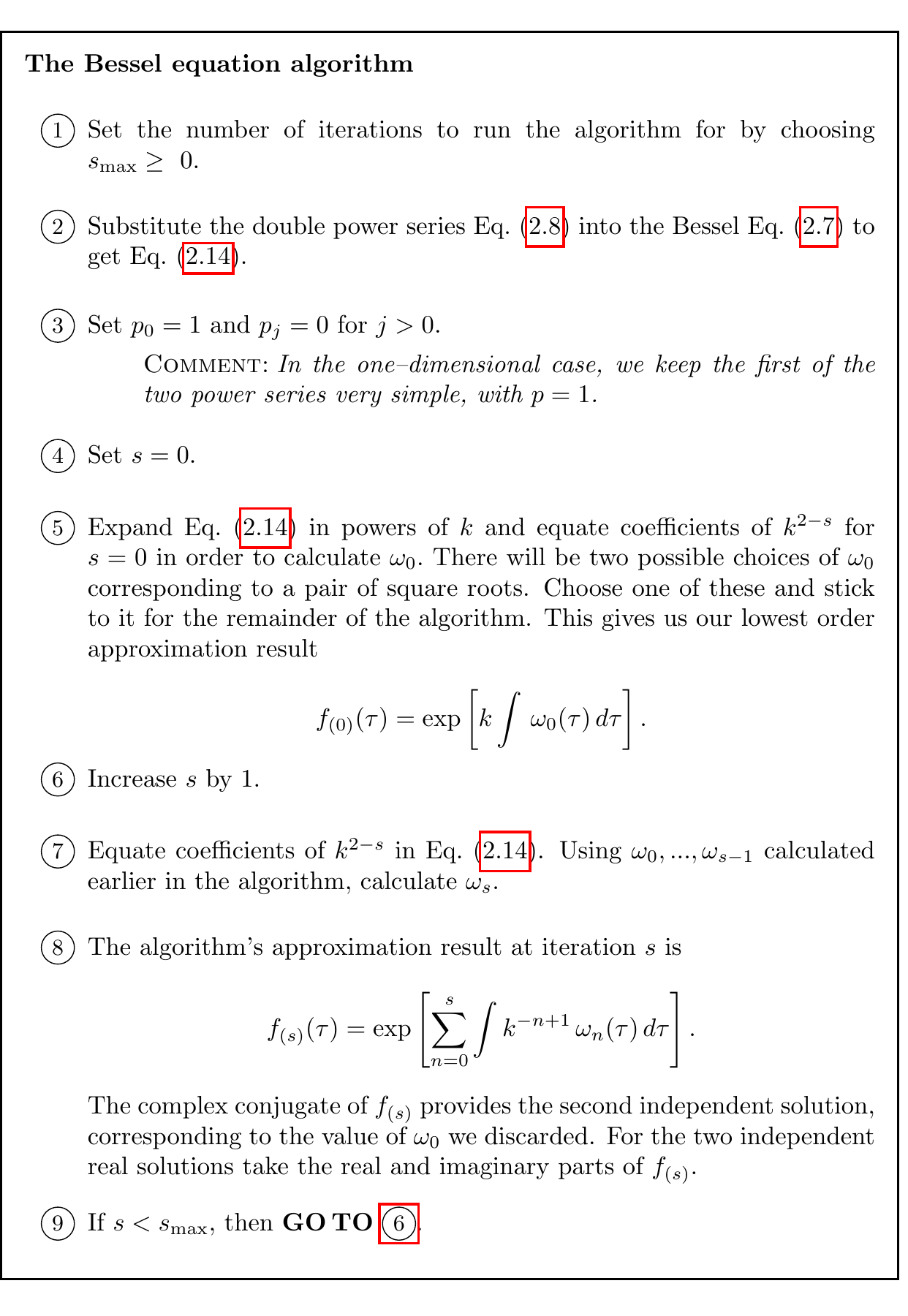}
	\caption[Bessel equation algorithm]{The double power series algorithm for approximate solutions to the Bessel Equation~\eqref{eq:besselexpn}}
	\label{fig:Besselalgorithm}
\end{figure}

From the calculations set out in Appendix~\ref{sec:Bessel-calculation},
we get the third order approximation to be
\be
\label{eq:Bessel-f3main}
f_{(3)}(\tn)= \frac{1}{\sqrt{\tn}}\exp\left[i\kn\tn-\frac{5i}{8\kn\tn}-\frac{5}{16\kn^2 \tn ^2}\right]
\,,
\ee
or its complex conjugate.  If we want real solutions, we take the real
and imaginary parts of \eref{eq:Bessel-f3main}, giving us the two real
independent solutions we expect for a second order differential
equation with real coefficients.

Note that, in this example, the $p$ power series of  Eqs.~\ref{eq:besselexpn} and \ref{eq:POdefn} is simply $p=1.$  Because we have two power series, $p$ and $\omega,$ we needed an additional  constraint to fully determine the terms $p_j$ and $\omega_n$ of \eref{eq:besselexpn}. To see this, note that when we take an order $\smax$ approximation, the terms $p_j$ for $j=0,1,...,\smax$ and $\omega_n$ for $n=0,1,...,\smax$ occur in our equations of coefficients. This means that we have $2\smax+2$ such terms to determine from the $\smax + 1$ equations coming from equating coefficients of $\kn^2,\kn,...,\kn^{2-\smax}.$  We therefore have to choose how to further constrain our choice of $p_j$ and $\omega_n$ terms in order to determine our approximation  fully. Since we are approximating oscillating functions, we want the $\omega$ series to do as much of the work as possible.  For a problem, such as this in the current example, in which there is only one underlying differential equation, we therefore simply set $p_0=1$ and all the other $p_j=0.$  If we choose different $p_j$ for $j>0,$ this makes the approximation expression more complicated, and, for plausible examples we have tested\footnote{Setting $p_j$ to be $\tn^{-j}$ times a random complex number with real and imaginary part each uniformly and independently drawn in the range $-1$ to $1.$}, makes virtually no difference to the accuracy of the approximation.  Essentially, the change in the $p$--terms is more or less balanced by consequent changes in the $\omega$--terms, in the range of $\tn$ for which the approximation is useful. 

This can be better understood looking more generally at cases, such as
for the cosmological perturbation problem in subsequent sections,
where we  have  more than one coupled differential equation
(Eqs.~\ref{eq:rgeq1c} and~\ref{eq:rgeq2c} for the cosmological
perturbation problem). When we have a number $d$ of coupled
differential equations, we see there are $d(\smax+1)$ of the $p_j$
terms and $(\smax+1)$ of the $\omega_n$ terms to be determined, giving
$(d+1)(\smax+1)$ terms in total.  These need to be determined from
$(\smax+1)$ coupled sets of $d$ equations, in other words from
$d(\smax+1)$ equations in total.  Looked at in terms of
\eref{eq:xexpn}, we have the $p_j$ now being $d$--dimensional vectors
$\mathbf{p}_j,$ which in general, for a given $\kn$ and $\tn,$ will be
constrained to lie in a $(d-1)$--dimensional hyperspace of
$\mathbb{C}^d.$ As discussed in
Appendix~\ref{sec:double-power-series-for-perturbations}, in the
paragraph preceding \eref{eq:ps}, we choose $\mathbf{p}_0$ to be a
suitable eigenvector associated with the coupled set of equations, and
this then constrains all but one degree of freedom~--- parallel to
that eigenvector~--- for each subsequent $\mathbf{p}_j.$ As suggested
in Ref.~\cite{FSN}, the remaining degree of freedom will be chosen by
setting so far undetermined terms equal to zero, in order to keep the
expression for the approximation as simple as possible.  As for the
case of a single differential equation, discussed following
\eref{eq:Bessel-f3main}, more complicated rules lead to more
complicated approximations of essentially the same accuracy. 
\\ 

The solution \eref{eq:Bessel-f3main} can be obtained easily
using our \emph{Mathematica} notebook, which we can also use to go to
higher orders.  For example, for the tenth order approximation, we get
\bml
\label{eq:f10soln}
f_{(10)}(\tn)=\frac{1}{\sqrt{\tn}}\exp\left[i \kn \tn-\frac{5 i}{8 \kn \tn}-\frac{5}{16 \kn^2 \tn^2}+\frac{145 i}{384 \kn^3 \tn^3}
\right.\\\left.
+\frac{85}{128 \kn^4 \tn^4}-\frac{1545 i}{1024 \kn^5 \tn^5}-\frac{25}{6
	\kn^6 \tn^6}
\right.\\\left.
+\frac{3097125 i}{229376 \kn^7 \tn^7}+\frac{205525}{4096 \kn^8 \tn^8}-\frac{165425425 i}{786432 \kn^9 \tn^9}\right]\,.
\eml\\

We want now to quantify the error in the approximation $f_{(s)},$ for each particular value of $s.$ One way to try to do this would be to take the analytical solution $f$ and calculate the ratio $\infrac{\left|\left(f_{(s)}-f\right)}{f}\right|.$  However, this runs into a difficulty.  Since $f$ is oscillating and repeatedly takes zero values, unless there is no error at all at these zeros, the ratio $\infrac{\left|\left(f_{(s)}-f\right)}{f}\right|$ will repeatedly become infinite. 
	
	We therefore adapt this approach in order to avoid this difficulty.  Broadly speaking, we estimate the typical error compared with the amplitude of oscillation.  To do this we first decide a target degree of accuracy, $1\%$ say.  We then draw a graph with a logarithmic scale to compare a plot of the error $\left|f_{(s)}-f\right|$ with a plot of the target accuracy, here $1\%\times \left|f\right|.$  Both plots will usually spike downwards towards zero repeatedly.  We consider that the approximation is accurate to within $1\%$ if, spikes when $f=0$ aside, $\left|f_{(s)}-f\right|\le 1\%\times \left|f\right|.$  The will show on the graph as the plot of $\left|f_{(s)}-f\right|$ being level with, or below, the plot of $1\%\times \left|f\right|$ (except near $f=0$ spikes).

\begin{figure}
	\centering
	\includegraphics[width=\linewidth]{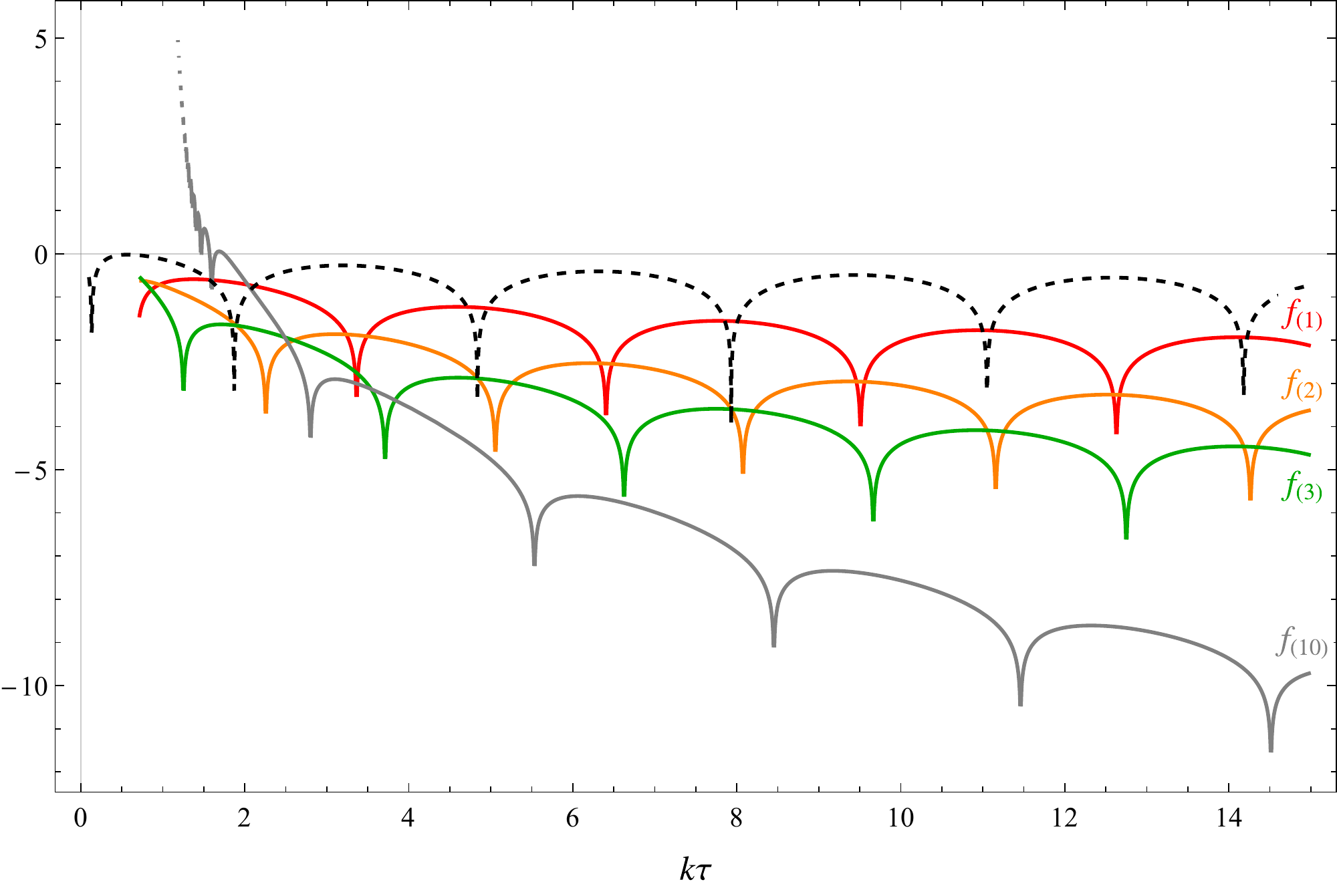}
	\caption[Bessel error plot]{The solid lines on this graph show the absolute errors in the approximations as ${\log}_{10}\left|\re\left(\infrac{f_{(s)}(\tn)-f(\tn)\right)}{\sqrt{\kn}}\right|,$ where $f$ is the exact solution. The function values are in units of $\sqrt{\kn}.$ For comparison, the dashed line shows 
		${\log}_{10}\left|\re\infrac{f(\tn)}{\sqrt{\kn}}\right|.$ The spikes in the lines represent periodic points where the numerical solution, or the absolute error, becomes zero.  A graph of the imaginary parts shows broadly similar features.}
	\label{fig:FigBesselerrorplot}
\end{figure}

Figure~\ref{fig:FigBesselerrorplot} follows this approach for $f_{(s)},$ with $s=1,2,3,10$. The error is small relative to the exact solution once $\kn\tn$ is sufficiently large. For larger $\kn\tn,$ the error associated with $f_{(s)}$ decreases as $s$ increases.  That rule does not necessarily apply for smaller $\tn$~--- for example, $f_{(10)}$ does not work as well as $f_{(3)}$ for $\kn\tn\lesssim 2.5.$  Section~\ref{sec:error-estimates} explores this issue further for the cosmological perturbation approximations which are the main subject of this paper.

We have shown how to use the double power series method to get good analytical approximate solutions to a second order differential equation of the form we are interested in in the following.  This approximation to the Bessel function is a new form of an approximation which can be found at 10.17.5 in \rfcite{NISTBessel}. Note that, in \rfcite{NISTBessel}, $J_i+iY_i$ is written as $H_i^{(1)}.$

\section{The perturbation equations} \label{sec:perturb}

We now study linear perturbations in a flat Friedmann--Robertson--Walker model with two non--interacting perfect fluids~--- radiation and dust (pressureless matter). The conformal time Friedmann equation for such a model is given by
\be\label{eq:Friedmann}
\H^2=\frac{8\pi G}{3} a^2 \left(\rhor+\rhom \right),
\ee
where $\H$ is the conformal Hubble factor, $a$ is the scale factor, $\rhor$ the homogeneous radiation and $\rhom$ the homogeneous matter density. As usual, the radiation and matter densities obey
\ba 
\rhor\propto a^{-4} & \text{\qquad and\qquad }  \rhom\propto a^{-3},
\ea
implying their rates of change with respect to conformal time are
\ba \label{eq:rho-rate}
\rhor'= -4\H \rhor & \text{\qquad and\qquad }  \rhom'= -3\H \rhom.
\ea

We focus on two second order differential equations, which contain only the metric potential and the radiation perturbations, and derive them from the standard system of perturbation equations. They will be useful to us as they form a system of the sort which is tractable with the double power series method. Once we solve those equations, an additional constraint equation gives the pressureless matter perturbation directly from the potential. 

As noted in the introduction when, in the following, we describe a solution as $n$th order, we mean that it is a double power series approximation to $n$th order of coefficients in the double power series.  This should not be confused with the order of the perturbation theory as described in, for example, \rfcite{2009PhR...475....1M}.

We also refer to \rfcite{2009PhR...475....1M} for the underlying perturbation equations and discussion of gauges. We use the longitudinal gauge, and study only linear order scalar perturbations. The line element is then given by
\be
ds^2=a^2\big(-(1+2\phi)d\td^2+(1-2\psi)\delta_{ij} dx^i dx^j\big)\,,
\ee
where $a$ is the scale factor and dependent only on the conformal time $\td,$ while $\phi, \psi$ and other perturbations are dependent on both $\td$ and the co--moving spatial co-ordinate $\vec{x}.$

In the longitudinal gauge, for perfect fluids and no anisotropic stresses, we have
\be 
\phi=\psi\,.
\ee
The trace of the spatial components of the Einstein equations gives the evolution equation
\be \label{eq:pertiinabla} 
\psi''+3\H\psi'+\left(2\H'+\H^2\right)\psi-4\pi G a^2 \delta P=0\,,
\ee
where a dash indicates a partial derivative with respect to the conformal time and $\delta P$ is the pressure perturbation.

We supplement this evolution equation with the energy and momentum conservation equations for a radiation fluid obeying the standard relationship between its pressure, $\Prr$ and its density, $\rhor,$
\be \label{eq:w}
\Prr=\frac{1}{3}\rho_\text{r}\,.
\ee
\steref{eq:w} relates the background radiation pressure to the background radiation density: the same proportionality relation holds between the radiation pressure and density perturbations.
From \rfcite{2009PhR...475....1M}, we have the energy and momentum conservation equations for radiation:
\bas\label{eq:enrpre}
0
&=
\delta\rhor'+3\H\left(\delta\rhor+\delta \Prr\right)-3(\rhor+\Prr)\psi'+(\rhor+\Prr)\nabla^2\Vr
\\
&=
\delta\rhor'+4\H\delta\rhor-4\rhor\psi'+\frac{4}{3}\rhor\nabla^2\Vr
\eas
and
\bas\label{eq:momrpre}
0
&=
\Vr'+\psi+\frac{\delta \Prr}{(\rhor+\Prr)}
\\
&=
\Vr'+\psi+\frac{\delta\rhor}{4\rhor}\,.
\eas

We now replace all our density perturbations, $\delta\rho_\alpha,$ with density contrast perturbations,
\be
\delta_\alpha=\frac{\delta\rho_\alpha}{\rho_\alpha}\,,
\ee
where $\alpha$ represents the fluid, here either radiation or pressureless matter.
We also replace the co--moving space co--ordinates $\vec{x}$ by their Fourier conjugates, the co--moving wave--numbers $\vec{\kd}.$  The Fourier transform substitutes $-\kd^2$ for $\nabla^2.$

In our radiation and pressureless matter model, the pressure perturbation in \eref{eq:pertiinabla} comes purely from radiation.  Written using density contrast perturbations, and Fourier transformed, that equation becomes
\be \label{eq:rgeq1c}
\psi''+3\H\psi'+\left(2\H'+\H^2\right)\psi=\frac{4\pi G}{3} a^2 \rhor\dr.
\ee
Using density contrast perturbations, and Fourier transformed, Eqs.~\eqref{eq:enrpre} and~\eqref{eq:momrpre} become
\be 
\dr'-4\psi'-\frac{4}{3} \kd^2\Vr=0
\quad\text{and}\quad
\Vr'+\psi+\frac{1}{4}\dr=0\,.
\ee
$\Vr$ can be eliminated between these two conservation equations giving
\be
\label{eq:rgeq2c}
4\psi''-\frac{4 \kd^2 }{3}\psi=\dr''+\frac{\kd^2}{3}\dr\,.
\ee
We will apply the double power series method to the pair of second differential equations, Eqs.~\eqref{eq:rgeq1c} and~\eqref{eq:rgeq2c}. Although the form of the \eref{eq:rgeq2c} may look a little unconventional, with two double derivatives, the method works without the need to diagonalise the equations~--- that is without the need to bring them to forms with only a single, and distinct, double derivative in each equation.

As noted above, we need a further equation to give us the matter perturbation in terms of $\psi.$  From~\rfcite{2009PhR...475....1M}, we have, as the $(0,0)$ component of the Einstein equations, the constraint equation
\be \label{eq:pert00nabla} 
3\H\psi'+3\H^2\psi+\kd^2\psi+4\pi G a^2 (\rhor \dr+\rhom\dm)=0\,,
\ee
where the $\dm$ is the fractional matter density perturbation and $\rhom$ the background matter density.

Note that there are \emph{four} independent solutions, or modes, in our system of equations. Equations~\eqref{eq:rgeq1c} and~\eqref{eq:rgeq2c} are a pair of simultaneous second order differential equations in two variables, $\dr$ and $\psi.$ They therefore have four modes.

\section{Leading order solutions}\label{sec:leading-order}

As mentioned in Section~\ref{sec:intro}, there are known analytical leading order approximations for perturbations in a flat radiation--matter model of the Universe.  Those for the oscillating modes arise from a form of the well known \emph{Wentzel-–Kramers-–Brillouin (WKB)} method for finding approximate analytical solutions to ordinary differential equations, see, for example,  Refs.~\cite{2006mmpe.book.....R,1978amms.book.....B}. WKB solutions have been known for a long time to be relevant for cosmological perturbation theory~--- see, for example, Refs.~\cite{1970ApJ...162..815P,2002ApJ...581..810W}. The solutions for the non--oscillating modes are different in character. They were originally derived, ignoring radiation perturbations completely, in \rfcite{1974A&A....37..225M}.

The leading order oscillating modes can be readily derived from Eqs. \eqref{eq:rgeq1c} and \eqref{eq:rgeq2c} of the previous section. To do this, we start by making an assumption~--- which we will verify below~--- that there are modes in which the partial derivative of potential perturbations with respect to the conformal time are of order $|\infrac{\partial\psi}{\partial \vec{x}}|\sim \kd\psi,$ and its double partial derivative with respect to conformal time is of order $|\infrac{\partial^2\psi}{\partial \vec{x}^2}|\sim \kd^2\psi,$ and we are working in the sub--horizon approximation, $\H\ll \kd.$ With these assumptions, \eref{eq:rgeq1c} gives us approximately
\be\label{eq:poissionish}
\psi''=\frac{4\pi G}{3} a^2\rhor\dr\,,
\ee
since $\H\psi'$ and $\H^2\psi$ can be neglected by the sub--horizon assumption. From the Friedmann equation \eref{eq:Friedmann}, we also have
\be
\frac{4\pi G}{3} a^2\rhor\le\frac{4\pi G}{3} a^2\left(\rhor+\rhom\right)=\frac{\H^2}{2}\,,
\ee
while, by assumption, we have that $\psi''$ is of order $\kd^2\psi.$ Putting these together, and neglecting factors of order unity, gives us that $\psi$ is of order at most $\infrac{\H^2\dr}{\kd^2}.$ Equation~\eqref{eq:rgeq2c} can then be approximated as
\be\label{eq:dr-approx}
0= \dr''-\frac{\kd^2}{3}\dr\,,
\ee
because the left--hand side of \eref{eq:rgeq2c}, having terms like $\kd^2\psi$ and $\psi'',$ is of order $\H^2\dr$ and so much less than its right--hand side, which is of order $\kd^2\dr.$ Solving \eref{eq:dr-approx}, this gives us the  leading order  WKB approximation for $\dr.$
\be\label{eq:wkb-r}
\dr(\td)=\exp\left[\frac{i\kd\td}{\sqrt{3}}\right]\,. 
\ee
To find the corresponding approximation for $\psi,$ note that the functions, $a$ and $\rhor$ on the right-hand of \eref{eq:poissionish} vary only slowly with respect to conformal time: $a'=\H a$ and, from \eref{eq:rho-rate}, $\rhor'=-4\H \rhor.$ Substitute \eref{eq:wkb-r} into \eref{eq:poissionish} and then solve \eref{eq:poissionish}, regarding $a$ and $\rhor$ as approximately constant, to get
\be\label{eq:wkb-psi}
\psi(\td)=-\frac{4\pi G  a^2\rhor}{\kd^2}\exp\left[\frac{i\kd\td}{\sqrt{3}}\right]\,.
\ee
This provides a solution  to \eref{eq:poissionish}, neglecting terms of order $\kd\H$ or $\H^2.$  We can also see this by substituting Eqs.~\ref{eq:dr-approx} and~\ref{eq:wkb-psi} into each of Eqs.~\ref{eq:rgeq1c} and~\ref{eq:rgeq2c} and noting that the approximations hold to zeroth order in $\H.$  \eref{eq:wkb-psi} also justifies our underlying order of magnitude assumptions about $\psi'$ and $\psi'',$   confirming the self--consistency of those assumptions.

 Using \eref{eq:pert00nabla}, we can also provide a corresponding approximation for the matter perturbation $\dm.$ The first try of this would be to retain only the leading order terms of \eref{eq:pert00nabla}, and substitute in the expressions for $\dr$ and $\psi$ from Eqs.~\eqref{eq:wkb-r} and~\eqref{eq:wkb-psi}.  However, this gives us 
\bml 
-\kd^2\frac{4\pi G  a^2\rhor}{\kd^2}\exp\left[\frac{i\kd\td}{\sqrt{3}}\right]
\\
+
4\pi G a^2 \left(\rhor\exp\left[\frac{i\kd\td}{\sqrt{3}}\right]+\rhom\dm\right)
=
0\,,
\eml
which would imply $\dm=0.$ We therefore seek the next order of approximation, which is given by also retaining the $3\H\psi'$ term from \eref{eq:pert00nabla}, which has one more factor of $\kd$ than the unused $3\H^2\psi$ term.  Using the $3\H\psi'$ term gives us
\bml \label{eq:wkb-m}
\dm
=
3i\H\frac{4\pi G  a^2\rhor}{4\pi G  a^2 \rhom \kd\sqrt{3}}\exp\left[\frac{i\kd\td}{\sqrt{3}}\right]
\\
=
\frac{\sqrt{3}i\H\arm}{\kd a}\exp\left[\frac{i\kd\td}{\sqrt{3}}\right]
\,.
\eml
 
The real and imaginary parts of Eqs.~\eqref{eq:wkb-r}, \eqref{eq:wkb-psi} and~\eqref{eq:wkb-m} provide the  leading order  WKB approximations for the two oscillating modes.
\\

Leading order approximations to the other two modes~--- the \Meszaros solutions~\cite{1974A&A....37..225M,1975A&A....41..143G}~--- can be derived in a similar fashion. $\dr$ can be eliminated between Eqs.~\eqref{eq:rgeq1c} and \eqref{eq:pert00nabla} giving
\bml \label{eq:matter}
\psi''+4\H\psi'+\left(2\H'+2\H^2+\frac{\kd^2}{3}\right)\psi
\\
=
-\frac{4\pi G}{3} a^2 \rhom\dm\,.
\eml
We can also obtain the energy and momentum conservation equations for matter from  \rfcite{2009PhR...475....1M}: they are
\be\label{eq:enmpre}
\dm'-3\psi'-\kd^2\Vm=0
\ee
and
\be\label{eq:mommpre}
\Vm'+\H\Vm=0\,. 
\ee
Eliminating $\Vm$ from these equations gives
\be\label{eq:m2}
\dm''+\H\dm'=3 \psi ''+3 \H \psi'-\kd^2 \psi\,.
\ee

We now outline the derivation of the \Meszaros solutions, starting by assuming $\psi'$ is of order $\H\psi$ and $\psi''$ of order $\H^2\psi.$ Equations~\eqref{eq:matter} and~\eqref{eq:m2} then give us the \Meszaros equation~\cite{1974A&A....37..225M}, 
\be \label{eq:Meszaroseqn}
\dm''+\H\dm'-4\pi G a^2\rhom\dm\approx 0\,.
\ee
The \Meszaros equation has two solutions, one growing and one decaying over time. \rfcite{1974A&A....37..225M} gives the growing solution to \eref{eq:Meszaroseqn} as being proportional to
\be \label{eq:Mes-grow}
\dm=1+\frac{3}{2}y\,,
\ee
where $y=\infrac{a}{\arm},$ where $\arm$ is the scale factor of radiation--matter equality~--- the moment when $\rhor=\rhom.$

The decaying solution of \eref{eq:Meszaroseqn} can be expressed in  analytical terms \cite{1975A&A....41..143G} as
\be \label{eq:Mes-decy}
\dm=\left(1+\frac{3}{2}y\right)\log\left[\frac{\sqrt{1+y}+1}{\sqrt{1+y}-1}\right]-3\sqrt{1+y}\,.
\ee
For either the growing or the decaying solution, we can also use Eqs.~\eqref{eq:matter} and~\eqref{eq:m2} to get
\be \label{eq:psimes}
\psi=-\frac{4\pi G}{\kd^2} a^2\rhom\dm
\ee
and  
\be\label{eq:Mes-end}
\dr=-4\psi\,.
\ee
 By substituting  these solutions back into  Eqs.~\ref{eq:rgeq1c} and~\ref{eq:rgeq2c}, we can check that these approximations, known as the \Meszaros solutions, are valid to zeroth order in $\H.$  Note that for each of these solutions, the $\psi$ and $\dr$ perturbations are suppressed by a factor of $\kd^2$ relative to the matter perturbation $\dm.$  They are approximate solutions to the full perturbation equations of Section~\ref{sec:perturb}: and they provide two of the four independent solutions. The other two solutions are the oscillating modes. 

\section{Solving the perturbed equations using the double power series method}\label{sec:prep}

We need to some preparation for using the double power series method.  For definiteness, we make explicit the dependence of terms in Eqs.~\eqref{eq:rgeq1c}, \eqref{eq:rgeq2c} and~\eqref{eq:pert00nabla} on the conformal time $\td$.  We also put the first resulting equations into matrix form.

\subsection{Background equations}

Using conformal time makes the $\kd^2$ term of \eref{eq:rgeq2c} slightly simpler (in cosmic time, it would instead be a $\infrac{\kd^2}{a^2}$ term).  Using conformal time also enables us to get a much simpler Friedmann solution for a flat radiation--matter universe than would be possible for cosmic time.

The relevant Friedmann equation was set out in \eref{eq:Friedmann}.
This can be written as
\be\label{eq:Friedmann2}
\H^2=\frac{\Hrm^2}{2}\left(\frac{\arm^4}{a^2}+\frac{\arm^3}{a} \right)\,,
\ee
where $\Hrm$ is the value of the conformal Hubble parameter at radiation--matter equality.

As noted in, for example, \rfcite{2006AIPC..843..111P}, integrating \eref{eq:Friedmann2} gives the conformal time expression for the scale factor 
\be \label{eq:rmscalefactor}
a(\td)=\arm\left(\frac{\td}{\tauc}+\frac{\td^2}{4\tauc^2}\right)\,,
\ee
where
\be \label{eq:taucrit}
\tauc=\frac{\sqrt{2}}{\Hrm}\,.
\ee
In \eref{eq:rmscalefactor}, the conformal time $\td$ is a non--negative number, with $a=0$ corresponding to $\td=0.$

 As mentioned near the start of Section~\ref{sec:introducing-the-double-power-series-method}, in the following  we will normalise the conformal time  to the dimensionless quantity $\tn=\infrac{\td}{\tauc}$ giving us
\be \label{eq:rmscale}
a(\tn)=\arm\left(\tn+\frac{\tn^2}{4}\right)\,.
\ee
The conformal time $\trm$ of radiation--matter equality is then found by setting $a(\tn)=\arm$ in \eref{eq:rmscale}, which gives us
\be \label{eq:rmscaletc}
\arm=\arm\left(\trm+\frac{\trm^2}{4}\right)\,,
\ee
for which the ($\tn>0$) solution is 
\be \label{eq:rmeq}
\trm=2(\sqrt{2}-1)\approx 0.828\,.
\ee

Using the cosmological parameters provided in \rfcite{2015arXiv150201589P} we find
\be 
\Hrm=\SI{0.0103}{a_0 Mpc^{-1}}\,.
\ee
The central estimate leads to
\be \label{eq:tauunit}
\tauc=\frac{\sqrt{2}}{\Hrm}=\SI{448}{a_0^{-1}.Myr}\,.
\ee
Recalling from \eref{eq:rmeq} that $\trm\approx0.828,$ we find a value in conventional units of $\td_\text{eq}\approx\SI{371}{a_0^{-1}.Myr}.$ 

 We normalise the co--moving wave--number, $\kd,$ on the same basis as we normalised $\td,$ setting $\kn=\kd\tauc.$ A value of $\kn=1$ is equivalent to a co--moving wave--number
\be \label{eq:kunit}
\tauc^{-1}=\frac{\Hrm}{\sqrt{2}}=\frac{a_0}{\SI{137}{Mpc}}\,,
\ee   
where, as is conventional, we have expressed the wave--number in units of inverse distance.
\\

From \eref{eq:rmscale}, the conformal Hubble parameter is given by
\be\label{eq:h-in-tau}
\H(\tn)=\frac{2\left(2+\tn\right)}{\tn\left(4+\tn\right)}
\ee
and its derivative is
\be\label{eq:hdash-in-tau}
\H'(\tn)=-\frac{2 \left(\tn ^2+4 \tn +8\right)}{\tn ^2 (\tn +4)^2}\,.
\ee
For the sub--horizon case we have $\kd\ll\H,$  corresponding to $(\kn\tn)^{-1}\ll 1$  and $\kn\gg\tn^{-1}.$  For a given co--moving wave--number, $\kn,$ the time $\tn_k$ such that
\be \label{eq:horizoncrossing}
\H(\tn_{\kn})=\frac{2\left(2+\tn_{\kn}\right)}{\tn_{\kn}\left(4+\tn_{\kn}\right)}=\kn
\ee
is known as the \emph{horizon--crossing time} for $\kn$. For the radiation--matter model, the ``horizon'' $\H$ is always shrinking with time, so we will sometimes refer to the crossing time as \emph{horizon entry}.

In Section~\ref{sec:an-approximate-solution-for-first-order-scalar-fast-modes-in-the-flat-radiation--matter-model}, we will use the example of $\kn=100$ to compare our approximate solutions with numerical results.  It is worth noting, from \eref{eq:tauunit}, that $\kn=100$ corresponds to a present day distance scale of \SI{1.37}{Mpc}. The (normalised) co--moving wavelength associated with $\kn$ is
\be 
\lambda=\frac{2\pi}{\kn}.
\ee
This implies the co--moving wavelength associated with $\kn=100$ is
\SI{8.64}{a_0^{-1}.Mpc}~--- at the present day, roughly the same
order of magnitude in size as clusters of galaxies (see
\rfcite{2013fgu..book.....L}, for example).  So $\kn=100$ perturbations
represent a scale similar to that of the largest gravitationally bound
structures in the present day Universe.

From \eref{eq:horizoncrossing}, we see that $\kn=100$ perturbations enter the horizon at $\tn\approx 0.01,$ when the scale factor is $a\approx 0.01\arm.$ This implies that $\kn=100$ perturbations enter the horizon  when $\infrac{\rhor}{\rhom}\approx 100,$ deep in the radiation--dominated epoch. 

It is also useful to note when radiation and baryons fully decouple. From that time onwards, until the much later time when dark energy becomes significant, our flat radiation--pressureless matter model reflects the make up of the actual Universe. \rfcite{2015arXiv150201589P} gives the relevant redshift~--- the end of the \emph{baryon drag} epoch~--- as $z_\text{dr}=\num{1060}.$ Note that, at this redshift,  radiation and matter are both significant constituents of the Universe's overall energy budget, with $23.9\%$ of the energy density being radiation.  We also note that the conformal time at $z_\text{dr}$ is $\tn_\text{dr}=\num{1.73}.$ 

\subsection{Perturbations}

We now express Eqs.~\eqref{eq:rgeq1c} and~\eqref{eq:rgeq2c} as matrix equations, along the lines of \eref{eq:matrxversion},

\be \label{eq:matrxversion2}
\matA\vecf''(\tn)+\matC\vecf'(\tn)+\matB\vecf(\tn)=\vec{0}\,.
\ee
We have
\ba \label{eq:fA}
\vecf &=
\begin{pmatrix}
\psi\\
\dr
\end{pmatrix},
& \matA&=
\begin{pmatrix}
1&0\\
-4&1
\end{pmatrix},
\ea
\ba
\matC=
\begin{pmatrix}
3\H&0\\
0&0
\end{pmatrix} 
\ea
and
\be \label{eq:fB}
\qquad 
\matB=
\begin{pmatrix}
2\H^2+2\H'&-\frac{4\pi G}{3}a^2 \rhor\\
\frac{4\kn^2}{3}&\frac{\kn^2}{3}
\end{pmatrix}.
\ee
We can write $\matB{}=\matB[-2]\,\kn^2 +\matB[0]$, where
\be \label{eq:B02a1}
\matB[-2]=\begin{pmatrix}
0&0\\
\frac{4}{3}&\frac{1}{3}
\end{pmatrix}
\ee
and
\be \label{eq:B02a2}
\matB[0] =\begin{pmatrix}
2\H^2+2\H'&-\frac{4\pi G}{3}a^2 \rhor \\
0&0
\end{pmatrix}.
\ee
Our subscripts on the $\matB$s correspond to the powers of $\kn^{-1}$ associated with the matrices in equation \eref{eq:matrxversion}.  As set out in \eref{eq:equaterule}, the negative index will assist us in keeping track of terms that arise in applying the double power series method.

We therefore have
\be
\label{eq:matrixdiff}
\matA\vecf''+ \matC\vecf'+\left(\matB[-2]\,\kn^2+\,\matB[0]\right) \vecf=\mathbf{0}\,,
\ee
with $\matA$, $\matB[-2]$, $\matB[0]$ and $\matC$ all independent of $\kn$, but depending on $\tn$.

We can use our convention that $\tunit=1$ to show that
\be \label{eq:rho-and-a1}
\frac{4\pi G}{3}a^2\rhor=\frac{1}{2}\left(\frac{\arm}{a}\right)^2
\ee
and
\be \label{eq:rho-and-a2}
\frac{4\pi G}{3}a^2\rhom=\frac{1}{2}\left(\frac{\arm}{a}\right)\,.
\ee
Using Eqs.~\ref{eq:rho-and-a1} and~ \ref{eq:rho-and-a2}, together with the expressions for $a$ from \eref{eq:rmscale} and for $\H$ from \eref{eq:h-in-tau}, we find that
\bas
\label{eq:B2Crm}
\matB[0]&=-\frac{8}{\tn^2\left(4+\tn\right)^2}
\begin{pmatrix}
2&1\\
0&0
\end{pmatrix}
\eas
and
\bas
\matC\label{eq:C-in-tn}
&=
\frac{6\left(2+\tn\right)}{\tn\left(4+\tn\right)}
\begin{pmatrix}
1&0\\
0&0
\end{pmatrix}.
\eas
This gives us the matrix equations explicitly in $\tn$ which we need to use the double power series method.

From \eref{eq:pert00nabla}, we find that, in terms of $\tn,$
\bml \label{eq:matter-tau}
\frac{6 (\tn +2)}{\tn  (4+\tn)}\psi'+ \frac{12 (\tn +2)^2}{\tn^2  (4+\tn)^2}\psi+\kn^2\psi
\\
+\frac{24}{\tn^2  (4+\tn)^2}
 \dr+\frac{6}{\tn  (4+\tn)}\dm=0\,.
\eml
Below we will use this to approximate the matter perturbation, $\dm.$

\section{Oscillating modes in a flat radiation--matter universe}\label{sec:an-approximate-solution-for-first-order-scalar-fast-modes-in-the-flat-radiation--matter-model}

We now apply the double power series method to the matrix equation of the previous section to
derive approximations to the oscillating modes. Our double power series method can be viewed as an extension of the  leading order  WKB approximation discussed in Section~\ref{sec:leading-order}.

\begin{figure} 
\centering
\includegraphics[width=\linewidth]{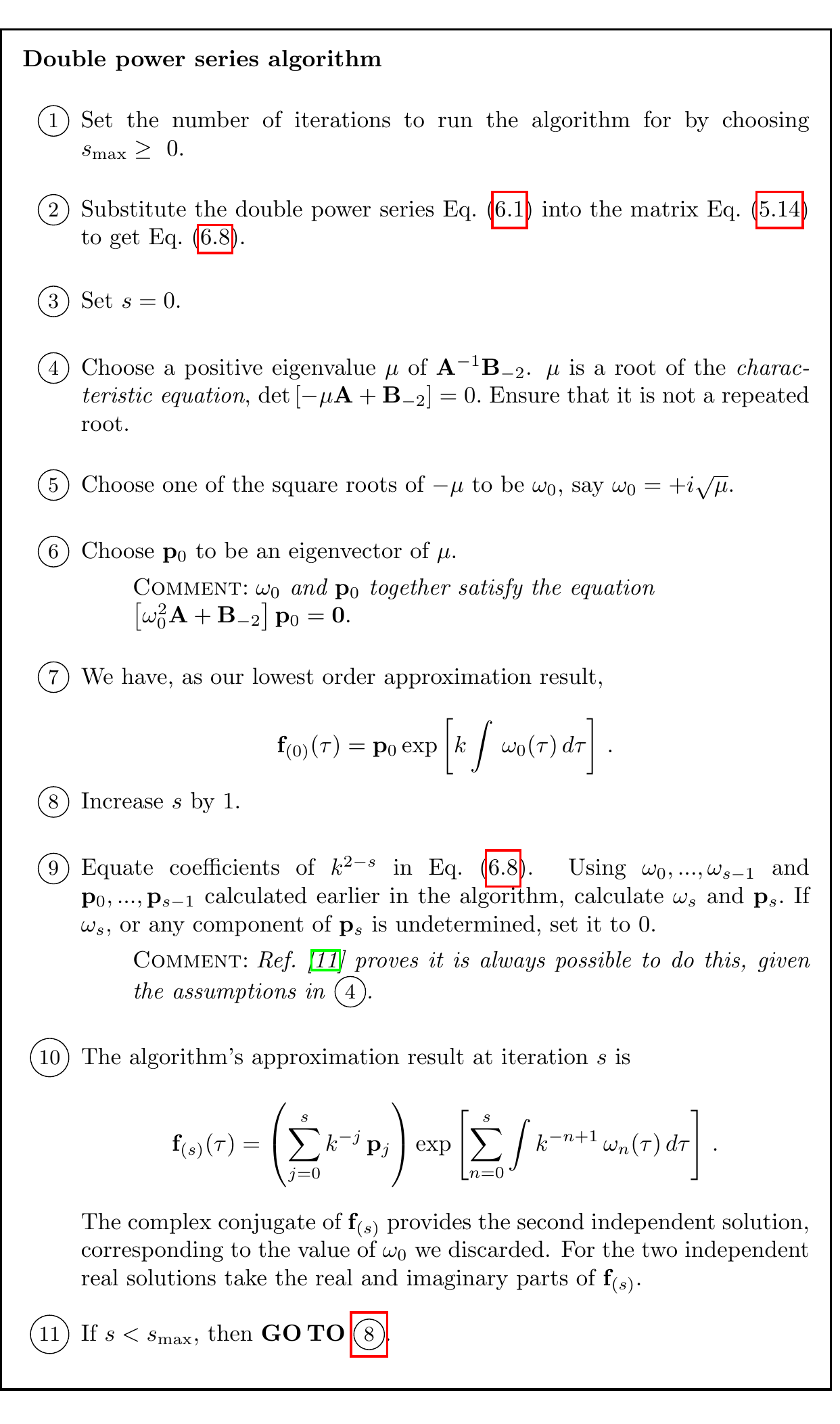}
\caption[The double power series algorithm]{The double power series algorithm for solving matrix differential equations of the form in \eref{eq:matrxversion2}.}
\label{fig:Doublepowerseries algorithm}
\end{figure}

The approach is analogous to that used in Section~\ref{sec:introducing-the-double-power-series-method} for the Bessel equation~--- see especially Figure~\ref{fig:Besselalgorithm}~--- but adapted to the matrix context, so that it works when we have system of differential equations, rather than just a single equation. The matrix algorithm is set out in Figure~\ref{fig:Doublepowerseries algorithm}. 

As defined in \eref{eq:xexpn}, the double power series method uses a series expansion
\bml \label{eq:vxexpn}
\vecf=
\begin{pmatrix}
\psi\\
\dr
\end{pmatrix}
\\
=\left(\sum_{j=0}^\infty \kn^{-j}\, \vec{p}_j\right) \exp\left[\sum_{n=0}^\infty \int \kn^{-n+1}\,\omega_n\,d\tn\right]\,,
\eml
where $\vecf$, $\vec{p}_j$ and $\omega_j$ are functions of $\tn$ and $\kn$ is regarded as a fixed parameter for $\vecf.$   The main difference from the Bessel function calculation is with regard to the $\vec{p}$ power series: we can now have $\vec{p}_j\ne 0$ for $j>0.$  

Following the Bessel example, we define
\begin{align} 
\vec{p}&=\sum_{j=0}^\infty \kn^{-j}\, \vec{p}_j,\label{eq:vPdefn}\\
\omega&=\sum_{n=0}^\infty \kn^{-n+1}\,\omega_n,\label{eq:vOdefn}
\end{align}
\begin{alignat}{2} \label{eq:vEdefn}
E&=\exp\left[\int\omega\,\mathrm{d}\tn\right]&=\exp\left[\sum_{n=0}^\infty \int \kn^{-n+1}\,\omega_n\,d\tn\right],
\end{alignat}
obtaining
\begin{align}
\vecf&=\vec{p} E,\label{eq:vf}\\
\vecf'&=\left(\vec{p}'+\vec{p}\,\omega\right)E\label{eq:vfd},\\
\mathrm{and\quad\quad} \vecf''&=\left(\vec{p}''+2\vec{p}'\,\omega+\vec{p}\,\omega'+\vec{p}\,\omega^2\right)E. \label{eq:vfdd}
\end{align}
Substituting these expressions into \eref{eq:matrxversion2}, we get
\bml \label{eq:doublepowerseriesapproxequation}
\matA\left(\vec{p}''+2\vec{p}'\,\omega+\vec{p}\,\omega'+\vec{p}\,\omega^2\right)+\matC\left(\vec{p}'+\vec{p}\,\omega\right)
\\
+\left(\matB[-2]\,\kn^2+\matB[0]\right)\vec{p}=\vec{0}\,.
\eml
We now apply the double power series algorithm, in the form set out in  Figure~\ref{fig:Doublepowerseries algorithm}.   As described  following \eref{eq:Besselapproxequation} for the Bessel equation example, we equate coefficients of powers of $\kn,$ starting with coefficients of $\kn^2,$ then using the results of that to equate coefficients of $\kn$ and so on, finally equating coefficients of $\kn^{2-\smax},$ where $\smax$ is the order to which we decide to take our approximation. We noted after \eref{eq:hdash-in-tau} that the sub—horizon assumption means $\kn\gg\tn^{-1}.$ The scale of the fast rate of change in the problem is set by $\kn,$ the wave--number\footnote{ We can express the fast rate of change more precisely as the wave--number $\kn$ multiplied by the magnitude of the relevant eigenvalue of the equation, as defined in Figure~\ref{fig:Doublepowerseries algorithm}. If that eigenvalue was very small or large, that would affect the fast rate of change of the problem.  However, in this case, the eigenvalue is $\infrac{i}{\sqrt{3}},$ giving the fast rate of change as $\infrac{\kn}{\sqrt{3}},$ which is itself of order $\kn.$}. At least for large enough $\kn$ and $\tn,$ the matrices $\matA,\matB[0]$ and $\matC,$ as set out in Eqs.~\eqref{eq:fA}, \eqref{eq:B2Crm} and~\eqref{eq:C-in-tn}, are also all of order much less than $\kn,$ and, relative to their size, they vary in (conformal) time only at the slow rate of change, of order $\tn^{-1}.$ This means that the leading order of an expression derived by our algorithm is always found simply by counting the powers of $\kn.$  This suggests we make our approximation by equating coefficients of powers of $\kn.$  

Appendix~\ref{sec:double-power-series-for-perturbations} works through derivation of the third order approximate solution for $\dr$ and $\psi.$  We can then get solutions for the matter perturbation $\dm$ from the constraint equation \eref{eq:matter-tau}. Note that in \eref{eq:matter-tau} some of the $\psi$ terms are multiplied by a factor of $\kn^2,$ which can come from the $\kn^2\psi$ term, but also, indirectly, from the $\psi''$ term when the double derivative acts on the exponential.  This means that to get a ``good'' approximation for $\dm$ to the same order as for $\psi$ and $\dr,$ we use a higher order solution for $\psi.$  We have used $\smax=3$ for our $\dr$ and $\psi,$ so here we use the $\psi$ approximation from $\smax=5$ to approximate $\dm$ to third order via \eref{eq:matter-tau}. The results for $\psi,\dr,$ and $\dm$ can also be calculated easily using a \emph{Mathematica} notebook at \rfcite{Github}.

Putting this together, we have that the third order approximate solution for the oscillating modes of linear scalar perturbations in a flat radiation--matter universe (assuming all matter is pressureless) are given by the real and imaginary parts of:
\begin{align} \label{eq:approx}
\begin{split}
\dr(\tn)&=\left(\frac{\tn }{4+\tn}\right)^{\frac{i \sqrt{3}}{\kn}}\exp\left[ \frac{i \kn \tn }{\sqrt{3}}+\frac{4 i \sqrt{3} (2+\tn)}{\kn \tn  (4+\tn)} \right],\\ \\
\psi(\tn)&=-\frac{24}{\kn^2 \tn ^2 (4+\tn)^2}\left\lbrace1-\frac{2 i\sqrt{3} \left( 2+\tn \right)}{\kn \tn (4+\tn)}\right\rbrace \dr(\tn),\\ \\
\dm(\tn)&=-\frac{144 }{\kn^2 \tn ^2 (4+\tn )^2} \left\lbrace1-\frac{5i \sqrt{3} (2+\tn)}{\kn \tn  (4+\tn)}
\right\rbrace \dr(\tn)\,.
\end{split}
\end{align}
Here $\tn$ is the value of the conformal time, when measured in units of $\tunit,$ and $\kn$ is the value of the co--moving wave--number, when measured in units of $\kunit.$

\begin{figure*}
\centering
\includegraphics[width=0.99\linewidth]{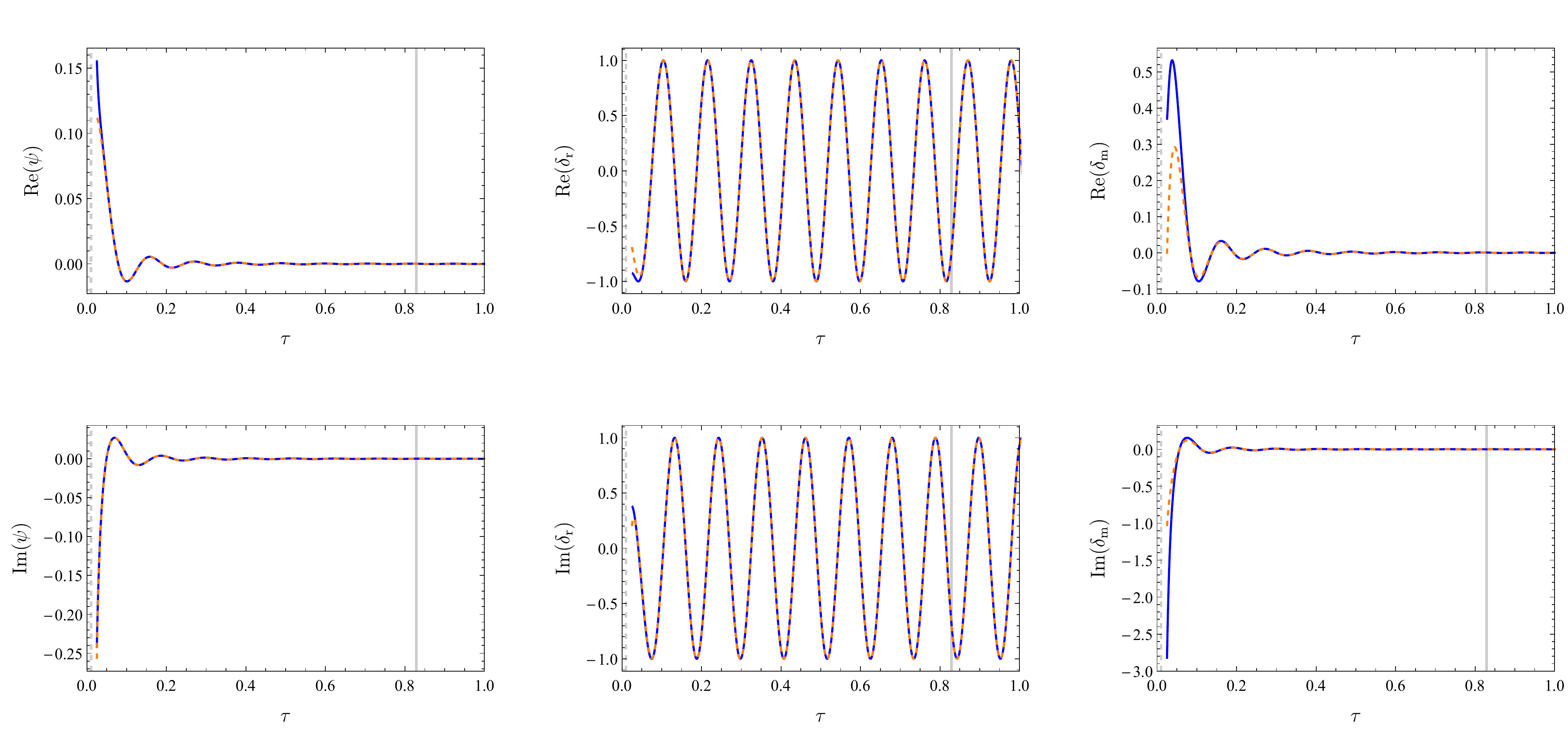}
\caption[Approximate solution for a flat radiation--matter universe, $\kn=100$]{The approximate solution~\eref{eq:approx} (solid blue line) and the numerical solution (dashed orange line) for Eqs.~\eqref{eq:rgeq1c} and \eqref{eq:rgeq2c} with $\kn=100$.  The columns show the real and imaginary parts respectively.  The horizontal axes show	$\tn.$ The solid vertical line corresponds to radiation--matter equality: see \eref{eq:rmeq}. The dashed vertical lines (for these graphs very close to the vertical axes) correspond to horizon crossing: see \eref{eq:horizoncrossing}. } 
\label{fig:Figrm100}

\centering
\includegraphics[width=0.99\linewidth]{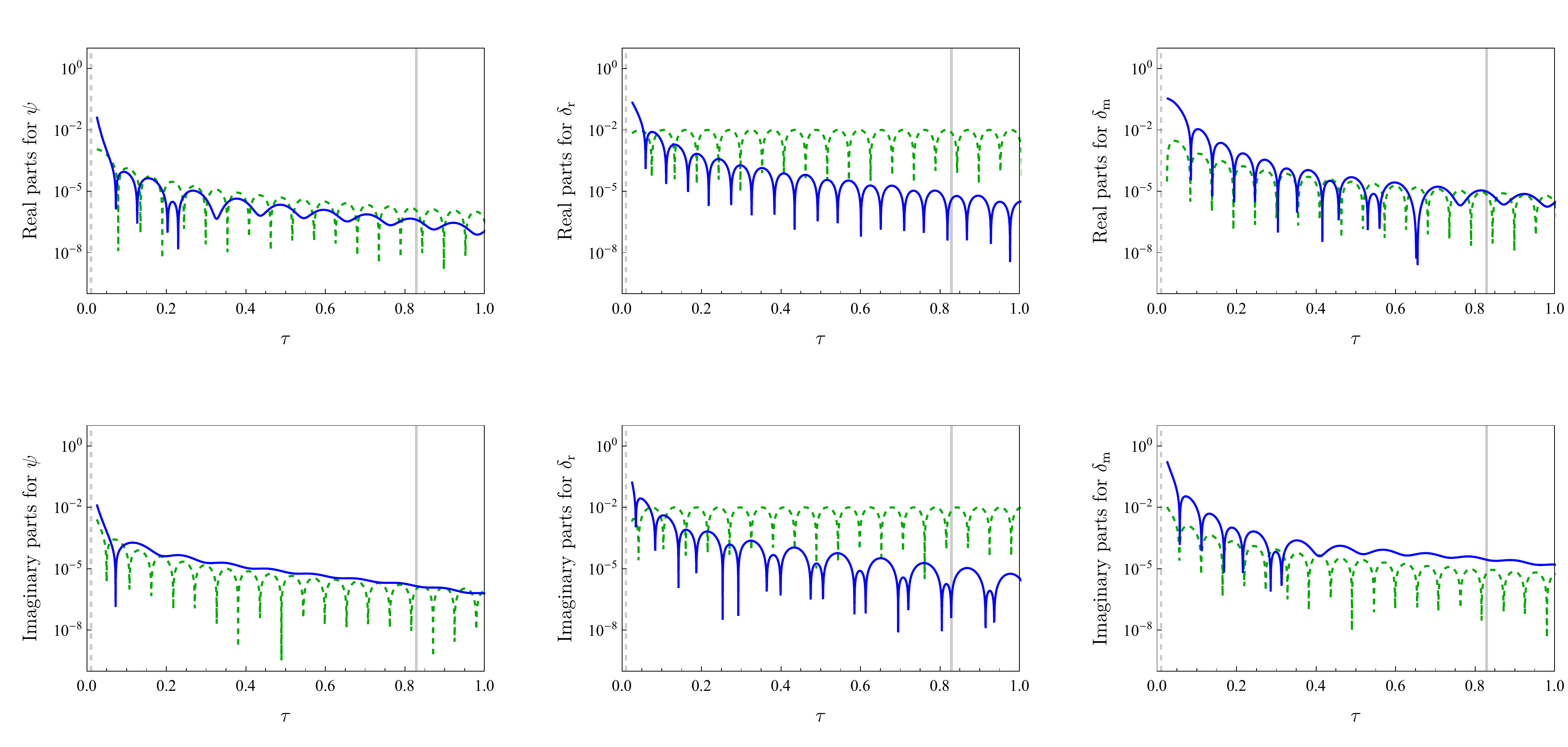}
\caption[Error plot for radiation--matter universe solutions, with $\kn=100$]{The heavy solid lines show the absolute errors in the radiation--matter universe $\kn=100$ approximations as ${\log}_{10}\left|f_{(3)}(\tn)-f(\tn)\right|,$ where $f_{(3)}$ is the relevant part of our approximation and $f$ is the relevant part of the numerical solution. ``Relevant'' means real or imaginary part of either $\psi$ or $\dr$ as indicated in the figure.  For comparison, the dashed line shows 
${\log}_{10}\left|1\%\times f(\tn)\right|.$ Downward spikes in the lines represent some points where the numerical solution, or the absolute error, becomes zero.}
\label{fig:Figrm100error}
\end{figure*}

We now plot this approximate solution for values of $\kn=100$ in Figure~\ref{fig:Figrm100}. As noted in Section~\ref{sec:prep}, such perturbations represent a scale similar to that of the largest gravitationally bound structures in the present day Universe. Figure~\ref{fig:Figrm100}  compares our approximate solutions with numerical solutions, derived using \emph{Mathematica}, for the perturbation equations. 

Figure~\ref{fig:Figrm100error} shows the errors in our approximate solutions, using a similar approach to that set out following \eref{eq:f10soln}.  For $\smax=3,$ we get a reasonable balance of having relatively few terms whilst being a valid approximation not too long after horizon--crossing and, once it is a valid approximation, having small errors, at least for $\psi$ and $\dr.$  Note that, because for some points the value of the exact numerical solution is $0,$ the percentage error is likely to grow very large at these period even if, in terms of absolute difference, the approximation is a good one.  Therefore, the best we can expect is for approximations to have a \emph{typical error} which is less than a given percentage. The typical error broadly corresponds to the error relative to the amplitude of oscillation.

\begin{figure*}
	\centering
	\includegraphics[width=0.99\linewidth 
	]{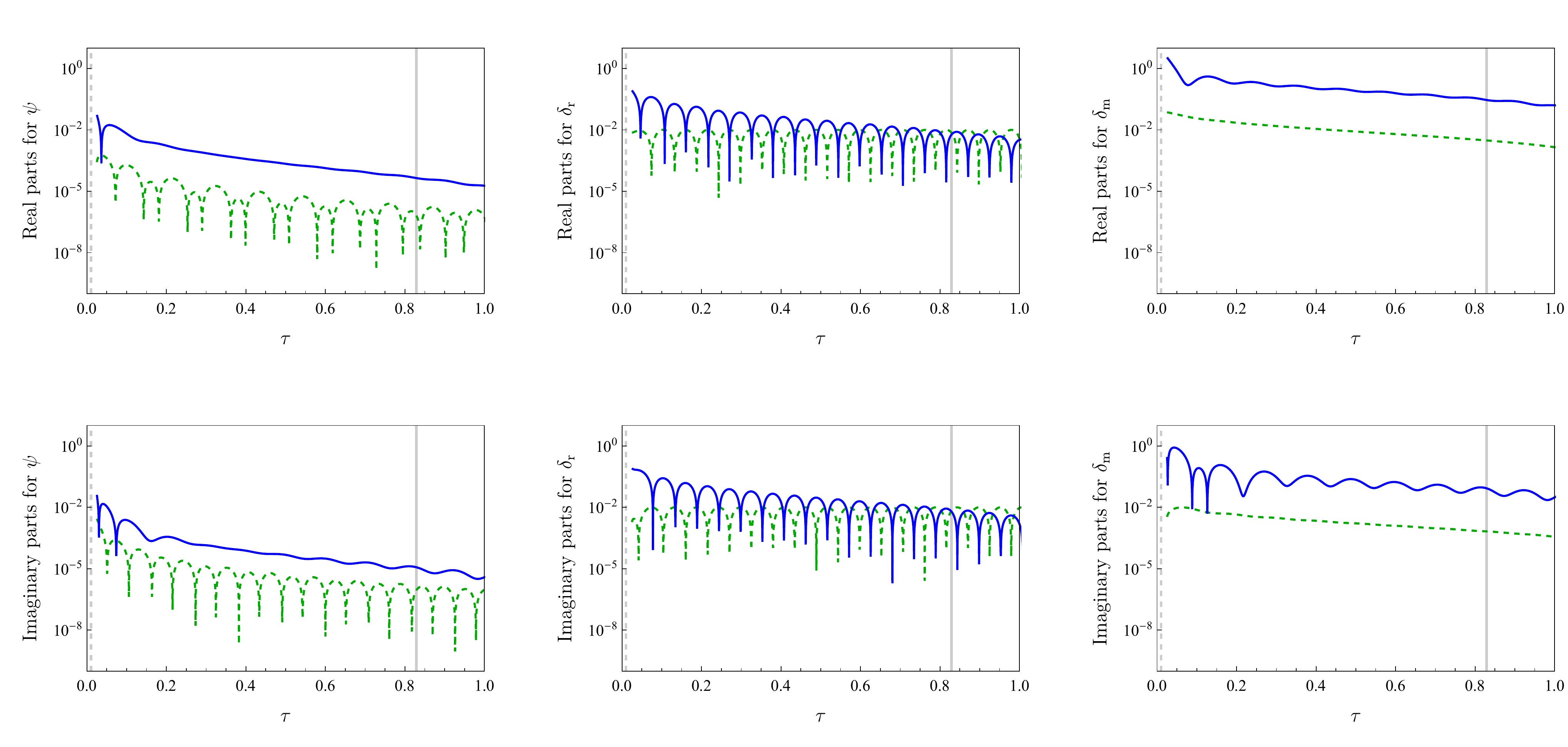}
	\caption[]{ As for Figure~\ref{fig:Figrm100error}, but for the  leading order  WKB approximation, Eqs.~\eqref{eq:wkb-r}, \eqref{eq:wkb-psi} and~\eqref{eq:wkb-m} of Section~\ref{sec:leading-order}. Note that we have matched boundary conditions for the $\dr$ approximation with those used in Figure~\ref{fig:Figrm100error}.  The $\dr$ WKB solution then determines the $\psi$ and $\dm$ WKB solutions, which therefore differ from those of Figure~\ref{fig:Figrm100error}, including at the boundary.}
	\label{fig:WKB100error}
\end{figure*}

Figure~\ref{fig:WKB100error} shows the error for the  leading order  WKB approximation of Eqs.~\eqref{eq:wkb-r}, \eqref{eq:wkb-psi} and~\eqref{eq:wkb-m} . Comparison with Figure~\ref{fig:Figrm100error} shows the improvement due to the double power series method.

\newpage \clearpage
Using Eqs.~\eqref{eq:rmscale} and~\eqref{eq:h-in-tau}, we can also write a solution proportional to that of \eref{eq:approx} as:
\bas\label{eq:happrox}
\dr(\td)&=\left(\frac{\td^2 }{a}\right)^{\frac{i \sqrt{3}}{\kd}}\exp\left[ \frac{i \kd \td }{\sqrt{3}}+\frac{2 i \sqrt{3}\H}{\kd} \right],\\
\psi(\td)&=-\frac{3}{4}\left(\frac{\Hrm}{\kd}\right)^2 \left(\frac{\arm}{a}\right)^2
\left\lbrace1-\frac{i\sqrt{3}\H}{\kd}\right\rbrace \dr(\td),\\
\dm(\td)&=-\frac{9}{2}\left(\frac{\Hrm}{\kd}\right)^2 \left(\frac{\arm}{a}\right)^2 \left\lbrace1-\frac{5i\sqrt{3}\H}{2\kd}
\right\rbrace \dr(\td)
\,,
\eas
 noting that in \eref{eq:happrox}  it is no longer necessary to require use of the normalised dimensionless versions of  the co--moving wave--number and conformal time,  because they always occur in contexts where the normalisation can be disregarded~--- such as $\kd\td$, $\infrac{\H}{\kd},$ or in determining the overall constant of proportionality for $\dr.$

We can improve our approximate solutions by going to higher orders. For example, the approximate seventh order solution $(\smax=7)$ is set out in Appendix~\ref{sec:app-seventh}, with the errors plotted. It provides a typical error for $\psi,\dr$ and $\dm$ of less than $1\%$ over a wide range of sub--horizon times and wave--numbers, including for $\kn=100$ and (plotted in the appendix's Figure~\ref{fig:Figrm10error}) $\kn=10.$ We chose this smaller value of $\kn$ to exhibit the seventh order solution because it appears that smaller $\kn$ tend to produce bigger errors than larger $\kn,$ so this test was more severe.

Using the double power series approximations, we now highlight some characteristics of the oscillating modes. From \eref{eq:happrox}, we note that the potential and matter perturbations for these modes are suppressed by a factor of $(\infrac{\Hrm\arm}{\kd a})^2$ relative to the radiation perturbations. Using \eref{eq:h-in-tau}~--- and recalling the choice of units in that equation~--- we can see that in the sub--horizon approximation this factor is always much less than $1.$

It can be shown that the absolute value of the larger eigenvalue of the $\matB$ matrix in \eref{eq:fB} is increasing with time (tending asymptotically to $\infrac{i\kn}{\sqrt{3}}$).  This suggests that the period of the oscillating modes should decrease with time.  We can see from our approximations that this inference is correct.  The period of oscillation may be represented by $\infrac{2\pi}{\omega},$ where $\omega$ is from \eref{eq:vOdefn}. From the third order approximation at \eref{eq:approx} or the seventh order approximation given in Appendix~\ref{sec:app-seventh}, we note that the oscillation period is asymptotically decreasing over time.  Take, for example, the seventh order approximation and $\kn=10.$ From $\tn=1$ to $\tn=10$ we see a decrease in period of $6\%$. We also note from Figure~\ref{fig:Figrm10error} that the typical error for $\dr$ within this range is much less than $1\%,$ confirming that this is a genuine effect.  This highlights the limitations of the  leading order~--- and hence constant period~---  WKB approximation.

Although not evident from the underlying perturbation equations, the seventh order approximation also suggests that the amplitude of the $\dr$ oscillations increases slightly over time.  The effect is small, around $0.2\%$ for $\kn=10$ over the range from $\tn=1$ to $\tn=10,$ with the variation nearly all occurring while $\tn<3.$ The typical error in that range is $0.1\%$ or less, and examination of the numerical solution also suggests that this effect, while small, is genuine.

\section{Error estimates}\label{sec:error-estimates}

In the previous section, we estimated the errors in our approximation by calculating numerical solutions and making comparisons.  We set out here a heuristic method for estimating errors without using numerical solutions.

\rfcite{1978amms.book.....B}, for example, notes that there is a heuristic method to estimate the errors in conventional power series approximations.  Consider such a series which approximates a function $g(x),$
\ba 
g(x)\sim \sum_{n=0}^\infty c_n (x-x_0)^n\,.
\ea
Then the heuristic rule is that the error from approximation by the $N$th partial sum,
\ba 
\left|g(x)-\sum_{n=0}^N c_n (x-x_0)^n\right|\,,
\ea
is approximately given by the absolute size of the $(N+1)$th term, $|c_{N+1}(x-x_0)^{N+1}|.$

For any given $x$, this rule also allows an estimate to be made as to which partial sum gives the best approximation to $g(x).$ If the series converges at $x,$ then the terms  $|c_{N+1}(x-x_0)^{N+1}|$ will eventually decrease indefinitely and each approximation will be an improvement on its predecessors.  However, if the series is divergent at $x,$ our heuristic rule for error estimation suggests we find the smallest term $|c_{N+1}(x-x_0)^{N+1}|$ for the given $x,$ and then
\ba 
\sum_{n=0}^N c_n (x-x_0)^n
\ea
gives the best approximation to $g(x).$ This will work if the terms, while not convergent, are sufficiently well--behaved. \rfcite{1978amms.book.....B} calls the best such approximation the \emph{optimal asymptotic expansion}.\\ 

For our double power series approximation we can proceed in a broadly similar fashion.  The approach is to start by focusing particularly on the approximation for $\dr,$ for which the relevant component of  the $\vecp$-series of \eref{eq:vxexpn} is equal to $1$.  We therefore focus on the $\omega$ power series,
\ba \label{eq:Omega-series}
\Omega={\sum_{n=0}^{\infty} \int \kn^{-n+1}\,\omega_n(\tn)\,d\tn}\,.
\ea
The $\dr$ component of the order $s$ approximation, $f_{(s)},$ is
\be \label{eq:dr-Omega}
\drs=\exp\left[{\sum_{n=0}^s \int \kn^{-n+1}\,\omega_n(\tn)\,d\tn}\right]
\equiv\exp\left[\Omega_{(s)}\right]\,.
\ee
The heuristic rule is that the corresponding error is estimated by the size of the $(s+1)$th term in the $\Omega$ series of \eref{eq:Omega-series},
\ba \label{eq:Omega-s}
\epsilon_s
&=
\left|\int \kn^{-(s+1)+1}\,\omega_{s+1}(\tn)\,d\tn\right|\,.
\ea

This also gives us a heuristic rule for identifying the best approximation.  For large enough $\tn,$ our series will be convergent and improve with each iteration.  For smaller $\tn,$ where the terms are still sufficiently well--behaved, we look for the smallest such term,
\ba 
\epsilon_{\smin}
&=
\left|\int \kn^{-(\smin+1)+1}\,\omega_{\smin+1}(\tn)\,d\tn\right|\,,
\ea 
and then take the corresponding order $\smin$ approximation to be the optimal approximation.  In some cases, the best approximation might be at $\smin-1$ or $\smin+1$, however, since we can assume that the error is at most of the same order as $\epsilon_{\smin},$ this need not be problematic.  The order $\smin$ double power series approximation also appears to give us close to optimal approximations for $\psi.$\\

We now test our heuristic method against the approach of comparison with numerical solutions used in Section~\ref{sec:an-approximate-solution-for-first-order-scalar-fast-modes-in-the-flat-radiation--matter-model} . Let $\drsnum$ be the numerical solution corresponding to the double power series approximation $\drs.$ As noted in \eref{eq:dr-Omega},  $\drs=\exp[{\Omega_{(s)}}]$. This suggests we use
\ba \label{eq:E-s}
E_{(s)}
=
\Bigg|\log_e\left[\frac{\drs}{\drsnum}\right]\Bigg|
=
\Bigg|\Omega_{(s)}-\log_e\left[\drsnum\right]\Bigg|
\ea
as our error estimate.

Table~\ref{table:error-estimates} sets out the results for $\kn=10,$ the value of $\kn$ used for the figures showing the seventh order approximation in Appendix~\ref{sec:app-seventh}.  In the table, $\tn$ ranges from $\tn=0.25$ to $\tn=1.$ Note the for $\tn=0.25,0.5$ and $0.75,$ our heuristic predicts the value of $s$ with the minimum error correctly.  For $\tn=1,$ that value of $s$ is out by $1$, but the difference in the accuracy of the approximation is minimal.

We have also looked at larger and smaller values of $\tn.$ For $\tn<0.25,$ we are approaching horizon--crossing and the series is ceasing to be well--behaved: the minimums of $\epsilon_s$ and $E_{(s)}$ need no longer closely correspond.  For $\tn>1,$ the ninth order approximation appears to remain the best approximation of below tenth order.

\begin{table*}
\bgroup
\begin{ruledtabular}
\def\arraystretch{1.5}
\addtolength{\jot}{9pt}
\caption[Error estimates]{Estimated and actual errors for $\dr$ with $\kn=10.$ $\epsilon_s$ and $E_{(s)}$ are defined in Eqs.~\ref{eq:Omega-s} and \ref{eq:E-s}. The smallest values in each of those columns is in bold. Notice that, for each value of $s$ and $\tn,$ we have $\epsilon_s \ge E_{(s)},$ as suggested by our heuristic method for error estimation.
\label{table:error-estimates}} 
\begin{tabular}{ddddddddddddd}
\tn & &\multicolumn{2}{c}{0.25}& & \multicolumn{2}{c}{0.5} & & \multicolumn{2}{c}{0.75} & & \multicolumn{2}{c}{1} \\
\hline
s & & \rcell{$\epsilon_s$\ \ \ \ }&  \rcell{$E_{(s)}$\ \ }& &\rcell{$\epsilon_s$\ \ \ \ }&  \rcell{$E_{(s)}$\ \ }& &\rcell{$\epsilon_s$\ \ \ \ }& \rcell{$E_{(s)}\ \ $}& &\rcell{$\epsilon_s$\ \ \ \ }&  \rcell{$E_{(s)}\ \ $} \\
 3 & & \bcell{0.3398} & \bcell{0.2819} & & 0.0498 & 0.0459 & & 0.0158 & 0.0092 & & 0.0068 & 0.0057 \\
  4 & & 0.6257 & 0.4757 & & 0.0458 & 0.0358 & & 0.0099 & 0.0078 & & 0.0033 & 0.0027 \\
  5 & & 0.7857 & 0.4341 & & \bcell{0.0357} & \bcell{0.0245} & & 0.0059 & 0.0045 & & 0.0016 & 0.0013 \\
  6 & & 1.2948 & 0.5310 & & 0.0373 & 0.0228 & & 0.0047 & 0.0033 & & 0.0010 & 0.0008 \\
  7 & & 3.1138 & 1.1122 & & 0.0507 & 0.0275 & & \bcell{0.0045} & \bcell{0.0030} & & 0.0008 & 0.0006 \\
  8 & & 9.6686 & 2.8491 & & 0.0843 & 0.0403 & & 0.0053 & 0.0032 & & \bcell{0.0007} & 0.0005 \\
  9 & & 0.6103 & 9.0526 & & 0.1685 & 0.0711 & & 0.0073 & 0.0040 & & 0.0008 & \bcell{0.0005} \\
 \end{tabular}
\end{ruledtabular}
\egroup
\end{table*}

\section{Conclusion}
\label{sec:conclusion}

We have used a double power series method which, to our
  knowledge, is new to cosmology to get sub--horizon approximations
for the oscillating modes of perturbations in a flat
radiation-pressureless matter universe.  Using \emph{Mathematica} or
similar packages it is quick and easy to calculate good higher order
approximations.

We approximated radiation and scalar potential perturbations in the
longitudinal gauge. From these, we then got the corresponding matter
perturbations.  The accuracy of approximation, when compared with
numerical solutions of the perturbation equations, will depend on the
wave--number and on the order to which the approximation is taken.  
Figures~\ref{fig:Figrm100error} and \ref{fig:Figrm10error} show that
the double power series approximation can achieve typical errors of
$1\%$ or better across a wide range of sub--horizon times for the
potential, radiation and matter perturbations.

Besides being interesting in themselves analytic solutions, derived
with the double power series method outlined in this work or
otherwise, have many applications. We hope in particular that they
will help to improve the efficiency of numerical calculations which
are computationally costly, if for example a large parameter space has
to be sampled. Using analytic solutions together with numerical ones
might increase the overall speed of the calculation without loosing
accuracy.
Analytical solutions can also be used as input for non--linear
governing equations. Recent examples using cosmological perturbation
theory at second order are the generation of vorticity
\cite{Christopherson:2009bt,Christopherson:2010ek}, and the generation
of magnetic fields \cite{Nalson:2013jya}. In both cases the evolution
equations for the second order quantities are sourced by the product
of two first order quantities. It was highlighted recently in
Ref.~\cite{Fidler:2015kkt}, that analytical methods might indeed be
more suitable for the study of magnetic field generation on small
scales than numerical schemes.

\vspace{10pt}
We will show in a forthcoming paper that the double power series
method can also be used to derive good approximations for tensor
perturbations in a flat radiation--matter universe. Another
  paper in preparation will show how to derive good approximations
for perturbations in a $\Lambda$CDM universe.  It is also possible
that the method may be useful in other models~--- for example, where
the acceleration of the late Universe is modelled using a quintessence
scalar field, or for modelling inflation.

The double power series method is also applicable for gauge choices other than the longitudinal gauge chosen in this paper, and for approximating linear perturbations in models of the universe with a single matter component, but where the equation of state varies over time.  For example, \rfcite{2013JCAP...01..002C} sets out the governing equation for a linear matter perturbation in the co--moving gauge, in a context where the equation of state varies over time.  Sub--horizon solutions to that equation can be analytically approximated by our method, providing the Hubble parameter and other time--dependent terms can be integrated analytically, either exactly or to a sufficiently good approximation. This illustrates the wide applicability of the double power series method in cosmology.

\vspace{10pt}
\begin{acknowledgments}
Some of this work was submitted by AJW to satisfy the requirements of
a MSc in Astrophysics at QMUL in 2015. KAM is supported, in part, by STFC grant ST/J001546/1.
\end{acknowledgments}

\bibliography{biblio}

\begin{thebibliography}{24}%
\makeatletter
\providecommand \@ifxundefined [1]{%
 \@ifx{#1\undefined}
}%
\providecommand \@ifnum [1]{%
 \ifnum #1\expandafter \@firstoftwo
 \else \expandafter \@secondoftwo
 \fi
}%
\providecommand \@ifx [1]{%
 \ifx #1\expandafter \@firstoftwo
 \else \expandafter \@secondoftwo
 \fi
}%
\providecommand \natexlab [1]{#1}%
\providecommand \enquote  [1]{``#1''}%
\providecommand \bibnamefont  [1]{#1}%
\providecommand \bibfnamefont [1]{#1}%
\providecommand \citenamefont [1]{#1}%
\providecommand \href@noop [0]{\@secondoftwo}%
\providecommand \href [0]{\begingroup \@sanitize@url \@href}%
\providecommand \@href[1]{\@@startlink{#1}\@@href}%
\providecommand \@@href[1]{\endgroup#1\@@endlink}%
\providecommand \@sanitize@url [0]{\catcode `\\12\catcode `\$12\catcode
  `\&12\catcode `\#12\catcode `\^12\catcode `\_12\catcode `\%12\relax}%
\providecommand \@@startlink[1]{}%
\providecommand \@@endlink[0]{}%
\providecommand \url  [0]{\begingroup\@sanitize@url \@url }%
\providecommand \@url [1]{\endgroup\@href {#1}{\urlprefix }}%
\providecommand \urlprefix  [0]{URL }%
\providecommand \Eprint [0]{\href }%
\providecommand \doibase [0]{http://dx.doi.org/}%
\providecommand \selectlanguage [0]{\@gobble}%
\providecommand \bibinfo  [0]{\@secondoftwo}%
\providecommand \bibfield  [0]{\@secondoftwo}%
\providecommand \translation [1]{[#1]}%
\providecommand \BibitemOpen [0]{}%
\providecommand \bibitemStop [0]{}%
\providecommand \bibitemNoStop [0]{.\EOS\space}%
\providecommand \EOS [0]{\spacefactor3000\relax}%
\providecommand \BibitemShut  [1]{\csname bibitem#1\endcsname}%
\let\auto@bib@innerbib\@empty
\bibitem [{\citenamefont {{Lifshitz}}(1946)}]{Lifshitz1946}%
  \BibitemOpen
  \bibfield  {author} {\bibinfo {author} {\bibfnamefont {E.~M.}\ \bibnamefont
  {{Lifshitz}}},\ }\bibfield  {title} {\enquote {\bibinfo {title} {{On the
  gravitational stability of the expanding universe}},}\ }\href
  {http://dx.doi.org/10.1007/s10714-016-2165-8} {\bibfield  {journal} {\bibinfo
   {journal} {Zhurnal Eksperimentalnoi i Teoreticheskoi Fiziki}\ }\textbf
  {\bibinfo {volume} {16}},\ \bibinfo {pages} {587--602} (\bibinfo {year}
  {1946})},\ \bibinfo {note} {as re-published in English translation in General
  Relativ Gravit \textbf{49} 18 (2017)}\BibitemShut {NoStop}%
\bibitem [{\citenamefont {{Peebles}}\ and\ \citenamefont
  {{Yu}}(1970)}]{1970ApJ...162..815P}%
  \BibitemOpen
  \bibfield  {author} {\bibinfo {author} {\bibfnamefont {P.~J.~E.}\
  \bibnamefont {{Peebles}}}\ and\ \bibinfo {author} {\bibfnamefont {J.~T.}\
  \bibnamefont {{Yu}}},\ }\bibfield  {title} {\enquote {\bibinfo {title}
  {{Primeval Adiabatic Perturbation in an Expanding Universe}},}\ }\href
  {\doibase 10.1086/150713} {\bibfield  {journal} {\bibinfo  {journal} {\apj}\
  }\textbf {\bibinfo {volume} {162}},\ \bibinfo {pages} {815} (\bibinfo {year}
  {1970})}\BibitemShut {NoStop}%
\bibitem [{\citenamefont {{Padmanabhan}}(2006)}]{2006AIPC..843..111P}%
  \BibitemOpen
  \bibfield  {author} {\bibinfo {author} {\bibfnamefont {T.}~\bibnamefont
  {{Padmanabhan}}},\ }\bibfield  {title} {\enquote {\bibinfo {title} {{Advanced
  Topics in Cosmology: A Pedagogical Introduction}},}\ }in\ \href {\doibase
  10.1063/1.2219327} {\emph {\bibinfo {booktitle} {Graduate School in
  Astronomy: X}}},\ \bibinfo {series} {American Institute of Physics Conference
  Series}, Vol.\ \bibinfo {volume} {843},\ \bibinfo {editor} {edited by\
  \bibinfo {editor} {\bibfnamefont {S.}~\bibnamefont {{Daflon}}}, \bibinfo
  {editor} {\bibfnamefont {J.}~\bibnamefont {{Alcaniz}}}, \bibinfo {editor}
  {\bibfnamefont {E.}~\bibnamefont {{Telles}}}, \ and\ \bibinfo {editor}
  {\bibfnamefont {R.}~\bibnamefont {{de la Reza}}}}\ (\bibinfo {year} {2006})\
  pp.\ \bibinfo {pages} {111--166},\ \Eprint
  {http://arxiv.org/abs/astro-ph/0602117} {arXiv:astro-ph/0602117} \BibitemShut
  {NoStop}%
\bibitem [{\citenamefont {{M\'esz\'aros}}(1974)}]{1974A&A....37..225M}%
  \BibitemOpen
  \bibfield  {author} {\bibinfo {author} {\bibfnamefont {P.}~\bibnamefont
  {{M\'esz\'aros}}},\ }\bibfield  {title} {\enquote {\bibinfo {title} {{The
  behaviour of point masses in an expanding cosmological substratum}},}\ }\href
  {http://articles.adsabs.harvard.edu/cgi-bin/nph-iarticle_query?1974A%26A....37..225M&amp;data_type=PDF_HIGH&amp;whole_paper=YES&amp;type=PRINTER&amp;filetype=.pdf}
  {\bibfield  {journal} {\bibinfo  {journal} {\aap}\ }\textbf {\bibinfo
  {volume} {37}},\ \bibinfo {pages} {225--228} (\bibinfo {year}
  {1974})}\BibitemShut {NoStop}%
\bibitem [{\citenamefont {{Groth}}\ and\ \citenamefont
  {{Peebles}}(1975)}]{1975A&A....41..143G}%
  \BibitemOpen
  \bibfield  {author} {\bibinfo {author} {\bibfnamefont {E.~J.}\ \bibnamefont
  {{Groth}}}\ and\ \bibinfo {author} {\bibfnamefont {P.~J.~E.}\ \bibnamefont
  {{Peebles}}},\ }\bibfield  {title} {\enquote {\bibinfo {title} {{Closed-form
  solutions for the evolution of density perturbations in some cosmological
  models}},}\ }\href
  {http://articles.adsabs.harvard.edu/cgi-bin/nph-iarticle_query?1975A%26A....41..143G&amp;data_type=PDF_HIGH&amp;whole_paper=YES&amp;type=PRINTER&amp;filetype=.pdf}
  {\bibfield  {journal} {\bibinfo  {journal} {\aap}\ }\textbf {\bibinfo
  {volume} {41}},\ \bibinfo {pages} {143--145} (\bibinfo {year}
  {1975})}\BibitemShut {NoStop}%
\bibitem [{\citenamefont {{Weinberg}}(2002)}]{2002ApJ...581..810W}%
  \BibitemOpen
  \bibfield  {author} {\bibinfo {author} {\bibfnamefont {S.}~\bibnamefont
  {{Weinberg}}},\ }\bibfield  {title} {\enquote {\bibinfo {title}
  {{Cosmological Fluctuations of Small Wavelength}},}\ }\href {\doibase
  10.1086/344441} {\bibfield  {journal} {\bibinfo  {journal} {\apj}\ }\textbf
  {\bibinfo {volume} {581}},\ \bibinfo {pages} {810--816} (\bibinfo {year}
  {2002})},\ \Eprint {http://arxiv.org/abs/astro-ph/0207375}
  {arXiv:astro-ph/0207375} \BibitemShut {NoStop}%
\bibitem [{\citenamefont {{Carroll}}\ \emph {et~al.}(1992)\citenamefont
  {{Carroll}}, \citenamefont {{Press}},\ and\ \citenamefont
  {{Turner}}}]{1992ARA&A..30..499C}%
  \BibitemOpen
  \bibfield  {author} {\bibinfo {author} {\bibfnamefont {S.~M.}\ \bibnamefont
  {{Carroll}}}, \bibinfo {author} {\bibfnamefont {W.~H.}\ \bibnamefont
  {{Press}}}, \ and\ \bibinfo {author} {\bibfnamefont {E.~L.}\ \bibnamefont
  {{Turner}}},\ }\bibfield  {title} {\enquote {\bibinfo {title} {{The
  cosmological constant}},}\ }\href {\doibase
  10.1146/annurev.aa.30.090192.002435} {\bibfield  {journal} {\bibinfo
  {journal} {\araa}\ }\textbf {\bibinfo {volume} {30}},\ \bibinfo {pages}
  {499--542} (\bibinfo {year} {1992})}\BibitemShut {NoStop}%
\bibitem [{\citenamefont {{Chernin}}\ \emph {et~al.}(2003)\citenamefont
  {{Chernin}}, \citenamefont {{Nagirner}},\ and\ \citenamefont
  {{Starikova}}}]{chernin2003growth}%
  \BibitemOpen
  \bibfield  {author} {\bibinfo {author} {\bibfnamefont {A.~D.}\ \bibnamefont
  {{Chernin}}}, \bibinfo {author} {\bibfnamefont {D.~I.}\ \bibnamefont
  {{Nagirner}}}, \ and\ \bibinfo {author} {\bibfnamefont {S.~V.}\ \bibnamefont
  {{Starikova}}},\ }\bibfield  {title} {\enquote {\bibinfo {title} {{Growth
  rate of cosmological perturbations in standard model: Explicit analytical
  solution}},}\ }\href {\doibase 10.1051/0004-6361:20021763} {\bibfield
  {journal} {\bibinfo  {journal} {\aap}\ }\textbf {\bibinfo {volume} {399}},\
  \bibinfo {pages} {19--21} (\bibinfo {year} {2003})},\ \Eprint
  {http://arxiv.org/abs/astro-ph/0110107} {arXiv:astro-ph/0110107} \BibitemShut
  {NoStop}%
\bibitem [{\citenamefont {Martin}\ and\ \citenamefont
  {Schwarz}(2003)}]{Martin:2002vn}%
  \BibitemOpen
  \bibfield  {author} {\bibinfo {author} {\bibfnamefont {Jerome}\ \bibnamefont
  {Martin}}\ and\ \bibinfo {author} {\bibfnamefont {Dominik~J.}\ \bibnamefont
  {Schwarz}},\ }\bibfield  {title} {\enquote {\bibinfo {title} {{WKB
  approximation for inflationary cosmological perturbations}},}\ }\href
  {\doibase 10.1103/PhysRevD.67.083512} {\bibfield  {journal} {\bibinfo
  {journal} {Phys. Rev.}\ }\textbf {\bibinfo {volume} {D67}},\ \bibinfo {pages}
  {083512} (\bibinfo {year} {2003})},\ \Eprint
  {http://arxiv.org/abs/astro-ph/0210090} {arXiv:astro-ph/0210090} \BibitemShut
  {NoStop}%
\bibitem [{\citenamefont {Casadio}\ \emph {et~al.}(2005)\citenamefont
  {Casadio}, \citenamefont {Finelli}, \citenamefont {Luzzi},\ and\
  \citenamefont {Venturi}}]{Casadio:2004ru}%
  \BibitemOpen
  \bibfield  {author} {\bibinfo {author} {\bibfnamefont {Roberto}\ \bibnamefont
  {Casadio}}, \bibinfo {author} {\bibfnamefont {Fabio}\ \bibnamefont
  {Finelli}}, \bibinfo {author} {\bibfnamefont {Mattia}\ \bibnamefont {Luzzi}},
  \ and\ \bibinfo {author} {\bibfnamefont {Giovanni}\ \bibnamefont {Venturi}},\
  }\bibfield  {title} {\enquote {\bibinfo {title} {{Improved WKB analysis of
  cosmological perturbations}},}\ }\href {\doibase 10.1103/PhysRevD.71.043517}
  {\bibfield  {journal} {\bibinfo  {journal} {Phys. Rev.}\ }\textbf {\bibinfo
  {volume} {D71}},\ \bibinfo {pages} {043517} (\bibinfo {year} {2005})},\
  \Eprint {http://arxiv.org/abs/gr-qc/0410092} {arXiv:gr-qc/0410092}
  \BibitemShut {NoStop}%
\bibitem [{\citenamefont {Feshchenko}\ \emph {et~al.}(1966)\citenamefont
  {Feshchenko}, \citenamefont {Shkil},\ and\ \citenamefont {Nikolenko}}]{FSN}%
  \BibitemOpen
  \bibfield  {author} {\bibinfo {author} {\bibfnamefont {S.~F.}\ \bibnamefont
  {Feshchenko}}, \bibinfo {author} {\bibfnamefont {N.~I.}\ \bibnamefont
  {Shkil}}, \ and\ \bibinfo {author} {\bibfnamefont {L.~D.}\ \bibnamefont
  {Nikolenko}},\ }\href {https://archive.org/details/nasa_techdoc_19670019838}
  {\emph {\bibinfo {title} {Asymptotic methods in the theory of linear
  differential equations, S.F. Feshchenko, N.I. Shkil, and L.D. Nikolenko.
  Translated by Scripta Technica}}}\ (\bibinfo  {publisher} {American Elsevier
  Pub. Co New York},\ \bibinfo {year} {1966})\ pp.\ \bibinfo {pages} {xvi, 270
  p.},\ \bibinfo {note} {translated for the National Aeronautics and Space
  Administration by John F. Holman and Co. Inc. NASw-1495 20546, available at
  https://archive.org/details/nasa\_ techdoc\_19670019838 online}\BibitemShut
  {NoStop}%
\bibitem [{\citenamefont {{Malik}}\ and\ \citenamefont
  {{Wands}}(2009)}]{2009PhR...475....1M}%
  \BibitemOpen
  \bibfield  {author} {\bibinfo {author} {\bibfnamefont {K.~A.}\ \bibnamefont
  {{Malik}}}\ and\ \bibinfo {author} {\bibfnamefont {D.}~\bibnamefont
  {{Wands}}},\ }\bibfield  {title} {\enquote {\bibinfo {title} {{Cosmological
  perturbations}},}\ }\href {\doibase 10.1016/j.physrep.2009.03.001} {\bibfield
   {journal} {\bibinfo  {journal} {\physrep}\ }\textbf {\bibinfo {volume}
  {475}},\ \bibinfo {pages} {1--51} (\bibinfo {year} {2009})},\ \Eprint
  {http://arxiv.org/abs/0809.4944} {arXiv:0809.4944 [astro-ph]} \BibitemShut
  {NoStop}%
\bibitem [{\citenamefont {{Wren}}\ and\ \citenamefont
  {{Malik}}(2016)}]{Github}%
  \BibitemOpen
  \bibfield  {author} {\bibinfo {author} {\bibfnamefont {A.~J.}\ \bibnamefont
  {{Wren}}}\ and\ \bibinfo {author} {\bibfnamefont {K.~A.}\ \bibnamefont
  {{Malik}}},\ }\href {https://github.com/AndrewWren/Double-power-series.git}
  {\enquote {\bibinfo {title} {{Mathematica notebooks for \emph{A double power
  series method for approximating cosmological perturbations}}},}\ } (\bibinfo
  {year} {2016}),\ \bibinfo {note} {{available at
  https://github.com/AndrewWren/Double-power-series.git. Note that, for
  deriving the numerical solutions used in Figs.~4-6 and~9-10, the standard
  \emph{Mathematica} numerical solver for differential equations requires the
  use of final rather than initial conditions for a stable solution. This is a
  purely numerical issue and nothing to do with our double power
  series.}}\BibitemShut {Stop}%
\bibitem [{\citenamefont {{Bender}}\ and\ \citenamefont
  {{Orszag}}(1999)}]{1978amms.book.....B}%
  \BibitemOpen
  \bibfield  {author} {\bibinfo {author} {\bibfnamefont {C.~M.}\ \bibnamefont
  {{Bender}}}\ and\ \bibinfo {author} {\bibfnamefont {S.~A.}\ \bibnamefont
  {{Orszag}}},\ }\href@noop {} {\emph {\bibinfo {title} {{Advanced Mathematical
  Methods for Scientists and Engineers: Asymptotic Methods and Perturbation
  Theory}}}},\ Vol.~\bibinfo {volume} {1}\ (\bibinfo  {publisher} {Springer},\
  \bibinfo {address} {New York},\ \bibinfo {year} {1999})\BibitemShut {NoStop}%
\bibitem [{{\relax DLMF}()}]{NISTBessel}%
  \BibitemOpen
  {\relax DLMF},\ \href {http://dlmf.nist.gov/10} {\enquote {\bibinfo {title}
  {{NIST Digital Library of Mathematical Functions}},}\ }\bibinfo
  {howpublished} {http://dlmf.nist.gov/, Release 1.0.9 of 2014-08-29} (\bibinfo
  {year} {2014}),\ \bibinfo {note} {online companion to
  \cite{Olver:2010:NHMF}}\BibitemShut {NoStop}%
\bibitem [{\citenamefont {{Riley}}\ \emph {et~al.}(2006)\citenamefont
  {{Riley}}, \citenamefont {{Hobson}},\ and\ \citenamefont
  {{Bence}}}]{2006mmpe.book.....R}%
  \BibitemOpen
  \bibfield  {author} {\bibinfo {author} {\bibfnamefont {K.~F.}\ \bibnamefont
  {{Riley}}}, \bibinfo {author} {\bibfnamefont {M.~P.}\ \bibnamefont
  {{Hobson}}}, \ and\ \bibinfo {author} {\bibfnamefont {S.~J.}\ \bibnamefont
  {{Bence}}},\ }\href@noop {} {\emph {\bibinfo {title} {{Mathematical Methods
  for Physics and Engineering - 3rd Edition}}}}\ (\bibinfo  {publisher}
  {Cambridge University Press},\ \bibinfo {address} {Cambridge},\ \bibinfo
  {year} {2006})\BibitemShut {NoStop}%
\bibitem [{\citenamefont {{Planck Collaboration}}\ \emph
  {et~al.}(2016)\citenamefont {{Planck Collaboration}}, \citenamefont {{Ade}},
  \citenamefont {{Aghanim}}, \citenamefont {{Arnaud}}, \citenamefont
  {{Ashdown}}, \citenamefont {{Aumont}}, \citenamefont {{Baccigalupi}},
  \citenamefont {{Banday}}, \citenamefont {{Barreiro}}, \citenamefont
  {{Bartlett}},\ and\ \citenamefont {et~al.}}]{2015arXiv150201589P}%
  \BibitemOpen
  \bibfield  {author} {\bibinfo {author} {\bibnamefont {{Planck
  Collaboration}}}, \bibinfo {author} {\bibfnamefont {P.~A.~R.}\ \bibnamefont
  {{Ade}}}, \bibinfo {author} {\bibfnamefont {N.}~\bibnamefont {{Aghanim}}},
  \bibinfo {author} {\bibfnamefont {M.}~\bibnamefont {{Arnaud}}}, \bibinfo
  {author} {\bibfnamefont {M.}~\bibnamefont {{Ashdown}}}, \bibinfo {author}
  {\bibfnamefont {J.}~\bibnamefont {{Aumont}}}, \bibinfo {author}
  {\bibfnamefont {C.}~\bibnamefont {{Baccigalupi}}}, \bibinfo {author}
  {\bibfnamefont {A.~J.}\ \bibnamefont {{Banday}}}, \bibinfo {author}
  {\bibfnamefont {R.~B.}\ \bibnamefont {{Barreiro}}}, \bibinfo {author}
  {\bibfnamefont {J.~G.}\ \bibnamefont {{Bartlett}}}, \ and\ \bibinfo {author}
  {\bibnamefont {et~al.}},\ }\bibfield  {title} {\enquote {\bibinfo {title}
  {{Planck 2015 results. XIII. Cosmological parameters}},}\ }\href {\doibase
  10.1051/0004-6361/201525830} {\bibfield  {journal} {\bibinfo  {journal}
  {\aap}\ }\textbf {\bibinfo {volume} {594}},\ \bibinfo {eid} {A13} (\bibinfo
  {year} {2016})},\ \Eprint {http://arxiv.org/abs/1502.01589} {arXiv:1502.01589
  [astro-ph.CO]} \BibitemShut {NoStop}%
\bibitem [{\citenamefont {{Loeb}}\ and\ \citenamefont
  {{Furlanetto}}(2013)}]{2013fgu..book.....L}%
  \BibitemOpen
  \bibfield  {author} {\bibinfo {author} {\bibfnamefont {A.}~\bibnamefont
  {{Loeb}}}\ and\ \bibinfo {author} {\bibfnamefont {S.~R.}\ \bibnamefont
  {{Furlanetto}}},\ }\href@noop {} {\emph {\bibinfo {title} {{The First
  Galaxies in the Universe}}}}\ (\bibinfo  {publisher} {Princeton University
  Press},\ \bibinfo {address} {Princeton},\ \bibinfo {year} {2013})\BibitemShut
  {NoStop}%
\bibitem [{\citenamefont {Christopherson}\ \emph {et~al.}(2009)\citenamefont
  {Christopherson}, \citenamefont {Malik},\ and\ \citenamefont
  {Matravers}}]{Christopherson:2009bt}%
  \BibitemOpen
  \bibfield  {author} {\bibinfo {author} {\bibfnamefont {Adam~J.}\ \bibnamefont
  {Christopherson}}, \bibinfo {author} {\bibfnamefont {Karim~A.}\ \bibnamefont
  {Malik}}, \ and\ \bibinfo {author} {\bibfnamefont {David~R.}\ \bibnamefont
  {Matravers}},\ }\bibfield  {title} {\enquote {\bibinfo {title} {{Vorticity
  generation at second order in cosmological perturbation theory}},}\ }\href
  {\doibase 10.1103/PhysRevD.79.123523} {\bibfield  {journal} {\bibinfo
  {journal} {Phys. Rev.}\ }\textbf {\bibinfo {volume} {D79}},\ \bibinfo {pages}
  {123523} (\bibinfo {year} {2009})},\ \Eprint {http://arxiv.org/abs/0904.0940}
  {arXiv:0904.0940 [astro-ph.CO]} \BibitemShut {NoStop}%
\bibitem [{\citenamefont {Christopherson}\ \emph {et~al.}(2011)\citenamefont
  {Christopherson}, \citenamefont {Malik},\ and\ \citenamefont
  {Matravers}}]{Christopherson:2010ek}%
  \BibitemOpen
  \bibfield  {author} {\bibinfo {author} {\bibfnamefont {Adam~J.}\ \bibnamefont
  {Christopherson}}, \bibinfo {author} {\bibfnamefont {Karim~A.}\ \bibnamefont
  {Malik}}, \ and\ \bibinfo {author} {\bibfnamefont {David~R.}\ \bibnamefont
  {Matravers}},\ }\bibfield  {title} {\enquote {\bibinfo {title} {{Estimating
  the amount of vorticity generated by cosmological perturbations in the early
  universe}},}\ }\href {\doibase 10.1103/PhysRevD.83.123512} {\bibfield
  {journal} {\bibinfo  {journal} {Phys. Rev.}\ }\textbf {\bibinfo {volume}
  {D83}},\ \bibinfo {pages} {123512} (\bibinfo {year} {2011})},\ \Eprint
  {http://arxiv.org/abs/1008.4866} {arXiv:1008.4866 [astro-ph.CO]} \BibitemShut
  {NoStop}%
\bibitem [{\citenamefont {Nalson}\ \emph {et~al.}(2014)\citenamefont {Nalson},
  \citenamefont {Christopherson},\ and\ \citenamefont
  {Malik}}]{Nalson:2013jya}%
  \BibitemOpen
  \bibfield  {author} {\bibinfo {author} {\bibfnamefont {Ellie}\ \bibnamefont
  {Nalson}}, \bibinfo {author} {\bibfnamefont {Adam~J.}\ \bibnamefont
  {Christopherson}}, \ and\ \bibinfo {author} {\bibfnamefont {Karim~A.}\
  \bibnamefont {Malik}},\ }\bibfield  {title} {\enquote {\bibinfo {title}
  {{Effects of non-linearities on magnetic field generation}},}\ }\href
  {\doibase 10.1088/1475-7516/2014/09/023} {\bibfield  {journal} {\bibinfo
  {journal} {JCAP}\ }\textbf {\bibinfo {volume} {1409}},\ \bibinfo {pages}
  {023} (\bibinfo {year} {2014})},\ \Eprint {http://arxiv.org/abs/1312.6504}
  {arXiv:1312.6504 [astro-ph.CO]} \BibitemShut {NoStop}%
\bibitem [{\citenamefont {{Fidler}}\ \emph {et~al.}(2016)\citenamefont
  {{Fidler}}, \citenamefont {{Pettinari}},\ and\ \citenamefont
  {{Pitrou}}}]{Fidler:2015kkt}%
  \BibitemOpen
  \bibfield  {author} {\bibinfo {author} {\bibfnamefont {C.}~\bibnamefont
  {{Fidler}}}, \bibinfo {author} {\bibfnamefont {G.}~\bibnamefont
  {{Pettinari}}}, \ and\ \bibinfo {author} {\bibfnamefont {C.}~\bibnamefont
  {{Pitrou}}},\ }\bibfield  {title} {\enquote {\bibinfo {title} {{Precise
  numerical estimation of the magnetic field generated around
  recombination}},}\ }\href {\doibase 10.1103/PhysRevD.93.103536} {\bibfield
  {journal} {\bibinfo  {journal} {\prd}\ }\textbf {\bibinfo {volume} {93}},\
  \bibinfo {eid} {103536} (\bibinfo {year} {2016})},\ \Eprint
  {http://arxiv.org/abs/1511.07801} {arXiv:1511.07801 [astro-ph.CO]}
  \BibitemShut {NoStop}%
\bibitem [{\citenamefont {{Christopherson}}\ \emph {et~al.}(2013)\citenamefont
  {{Christopherson}}, \citenamefont {{Hidalgo}},\ and\ \citenamefont
  {{Malik}}}]{2013JCAP...01..002C}%
  \BibitemOpen
  \bibfield  {author} {\bibinfo {author} {\bibfnamefont {A.~J.}\ \bibnamefont
  {{Christopherson}}}, \bibinfo {author} {\bibfnamefont {J.~C.}\ \bibnamefont
  {{Hidalgo}}}, \ and\ \bibinfo {author} {\bibfnamefont {K.~A.}\ \bibnamefont
  {{Malik}}},\ }\bibfield  {title} {\enquote {\bibinfo {title} {{Modelling
  non-dust fluids in cosmology}},}\ }\href {\doibase
  10.1088/1475-7516/2013/01/002} {\bibfield  {journal} {\bibinfo  {journal}
  {JCAP}\ }\textbf {\bibinfo {volume} {1}},\ \bibinfo {eid} {002} (\bibinfo
  {year} {2013})},\ \Eprint {http://arxiv.org/abs/1207.1870} {arXiv:1207.1870
  [astro-ph.CO]} \BibitemShut {NoStop}%
\bibitem [{\citenamefont {Olver}\ \emph {et~al.}(2010)\citenamefont {Olver},
  \citenamefont {Lozier}, \citenamefont {Boisvert},\ and\ \citenamefont
  {Clark}}]{Olver:2010:NHMF}%
  \BibitemOpen
  \bibinfo {editor} {\bibfnamefont {F.~W.~J.}\ \bibnamefont {Olver}}, \bibinfo
  {editor} {\bibfnamefont {D.~W.}\ \bibnamefont {Lozier}}, \bibinfo {editor}
  {\bibfnamefont {R.~F.}\ \bibnamefont {Boisvert}}, \ and\ \bibinfo {editor}
  {\bibfnamefont {C.~W.}\ \bibnamefont {Clark}},\ eds.,\ \href@noop {} {\emph
  {\bibinfo {title} {{NIST Handbook of Mathematical Functions}}}}\ (\bibinfo
  {publisher} {Cambridge University Press},\ \bibinfo {address} {New York,
  NY},\ \bibinfo {year} {2010})\ \bibinfo {note} {print companion to
  \cite{NISTBessel}}\BibitemShut {NoStop}%
\end{thebibliography}%


\appendix

\section{The double power series approximation for a Bessel function}
\label{sec:Bessel-calculation}

In this appendix, we work out explicitly Section~\ref{sec:introducing-the-double-power-series-method}'s Bessel function solution  to third order, following the algorithm in Figure~\ref{fig:Besselalgorithm}.  Circled numbers refer to the steps of that algorithm.\\ \\
\vspace{6 pt}\noindent\textbf{\nc{1}}\nopagebreak \\ \nopagebreak
In this example, we take $\smax=3.$\\ \\ 
\vspace{6 pt}\textbf{\nc{2} and \nc{3}}\nopagebreak \\ \nopagebreak
Taking these two steps together, we put $p=1$ into \eref{eq:Besselapproxequation} to get
\be \label{eq:Besselapproxequation2}
\left(\,\omega'+\,\omega^2\right)+\frac{1}{\tn}\,\omega+\left(\kn^2+\frac{1}{\tn^2}\right)=0.
\ee \\ 
\vspace{6 pt}\noindent \textbf{\nc{4} and \nc{5}}\nopagebreak \\ \nopagebreak
Set $\underline{s=0}$. Coefficients of $\kn^{2-0}=\kn^2$ in \eref{eq:Besselapproxequation2} can only come from its $\omega^2$ and $\kn^2$ terms.  Equating coefficients of $\kn^2,$ we get
\be
\omega_0^2+1=0. 
\ee
So
\be \label{eq:o0}
\omega_0=\pm i
\ee
and we choose one of the solutions, say $\omega_0=+i$. From Figure~\ref{fig:Besselalgorithm}, our lowest order approximation is therefore
\be 
f_{(0)}(\tn)= \exp\left[\int i\kn d\tn\right] = \exp\left[i\kn\tn\right].
\ee
We choose the constant of integration to be $0$: any other choice would simply multiply $f_{(0)}$ by a constant.

\begin{figure}
\centering
\includegraphics[width=\linewidth]{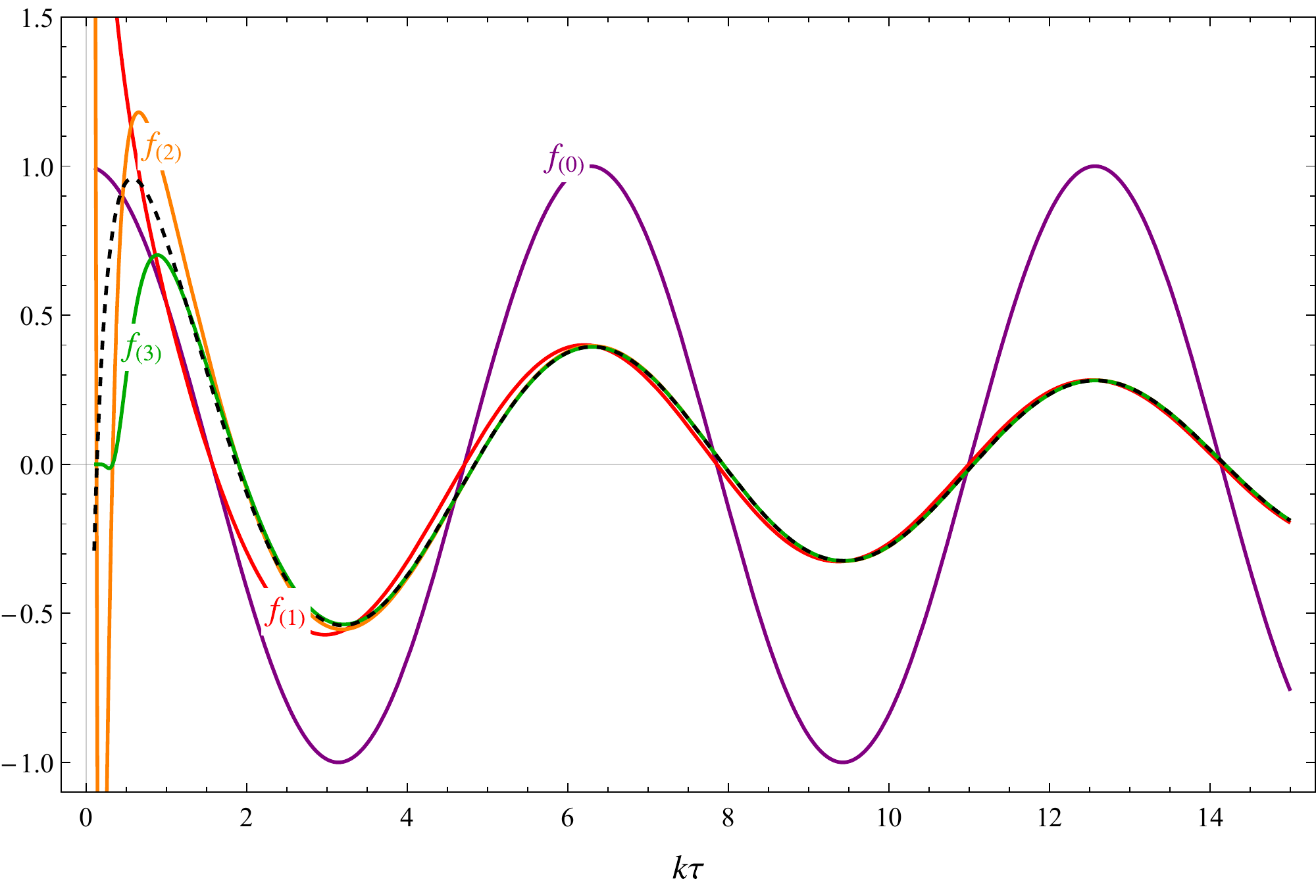}
\caption[Approximate and exact Bessel solutions]{This figure shows real parts of the approximate (solid lines) and exact (dashed line) analytical solutions to \eref{eq:bessel-dls}. The vertical axis is in units of $\sqrt{\kn}.$ The calculation of the exact solution is explained in Figure~\ref{fig:matchingbessel}.  A graph of the imaginary parts shows broadly similar features.}
\label{fig:FigBesselcompare}
\end{figure}

A comparison of this solution with the exact analytical solution given by Bessel functions can be found in Figure~\ref{fig:FigBesselcompare}~--- see the purple solid line and the black dashed line respectively.  We note that $f_{(0)}$ does not match the analytical solution very well~--- in particular, it does not have the latter's decaying amplitude.\\ \\
\\
\vspace{6 pt}\noindent\textbf{\nc{6} to \nc{9}}\nopagebreak \\ \nopagebreak 
We increase $s$ by $1$ to get $\underline{s=1}.$ We now equate coefficients of $\kn^{2-1}=\kn^1$ in \eref{eq:Besselapproxequation2}, making use of $\omega_0=i$. This gives us
\be
\omega_0'+2\omega_0\omega_1 +\frac{1}{\tn}\omega_0 =2i\omega_1 +\frac{1}{\tn}i =0,
\ee
yielding
\be
\omega_1=-\frac{1}{2\tn}.
\ee
This gives
\bas
f_{(1)}(\tn)&= \exp\left[\int \left(i\kn-\frac{1}{2\tn}\right)d\tn\right]
\\
&=
\exp\left[i\kn\tn-\frac{1}{2}\log(\tn)\right] \\
&= \frac{1}{\sqrt{\tn}}\exp\left[i\kn\tn\right].
\eas
As elsewhere in this paper, unless otherwise specified, all logarithms are natural logarithms to the base $e$. Figure~\ref{fig:FigBesselcompare}'s red line shows that this is a much better solution than $f_{(0)},$ being very close to the analytical solution for $\kn\tn\gtrsim 6.$

We now increase $s$ by $1,$ to get $\underline{s=2}.$ Equating coefficients of $\kn^{2-2}=1$ in \eref{eq:Besselapproxequation2}, gives us
\be
\left(\omega_1'+\left[2\omega_0\omega_2+\omega_1^2\right]\right)+\frac{1}{\tn}\,\omega_1+\frac{1}{\tn^2}=0.
\ee
Using the values of $\omega_0$ and $\omega_1$ calculated above, we have
\be
\left(\frac{1}{2\tn^2}+\left[2i\omega_2+\frac{1}{4\tn^2}\right]\right)-\frac{1}{2\tn^2}+\frac{1}{\tn^2}=0,
\ee
giving
\be
\omega_2=\frac{5i}{8\tn^2}.
\ee
This gives us
\bas
f_{(2)}(\tn)&= \exp\left[\int \left(i\kn-\frac{1}{2\tn}+\frac{5i}{8\kn\tn^2}\right)d\tn\right]\\
&=\exp\left[i\kn\tn-\frac{1}{2}\log(\tn)-\frac{5i}{8\kn\tn}\right]\\
&= \frac{1}{\sqrt{\tn}}\exp\left[i\kn\tn-\frac{5i}{8\kn\tn}\right].
\eas
Figure~\ref{fig:FigBesselcompare}'s orange line shows that this is a further improvement: $f_{(2)}$ is very close to the analytical solution for $\kn\tn\gtrsim 3.5.$

\begin{figure}
\centering
\includegraphics[width=\linewidth]{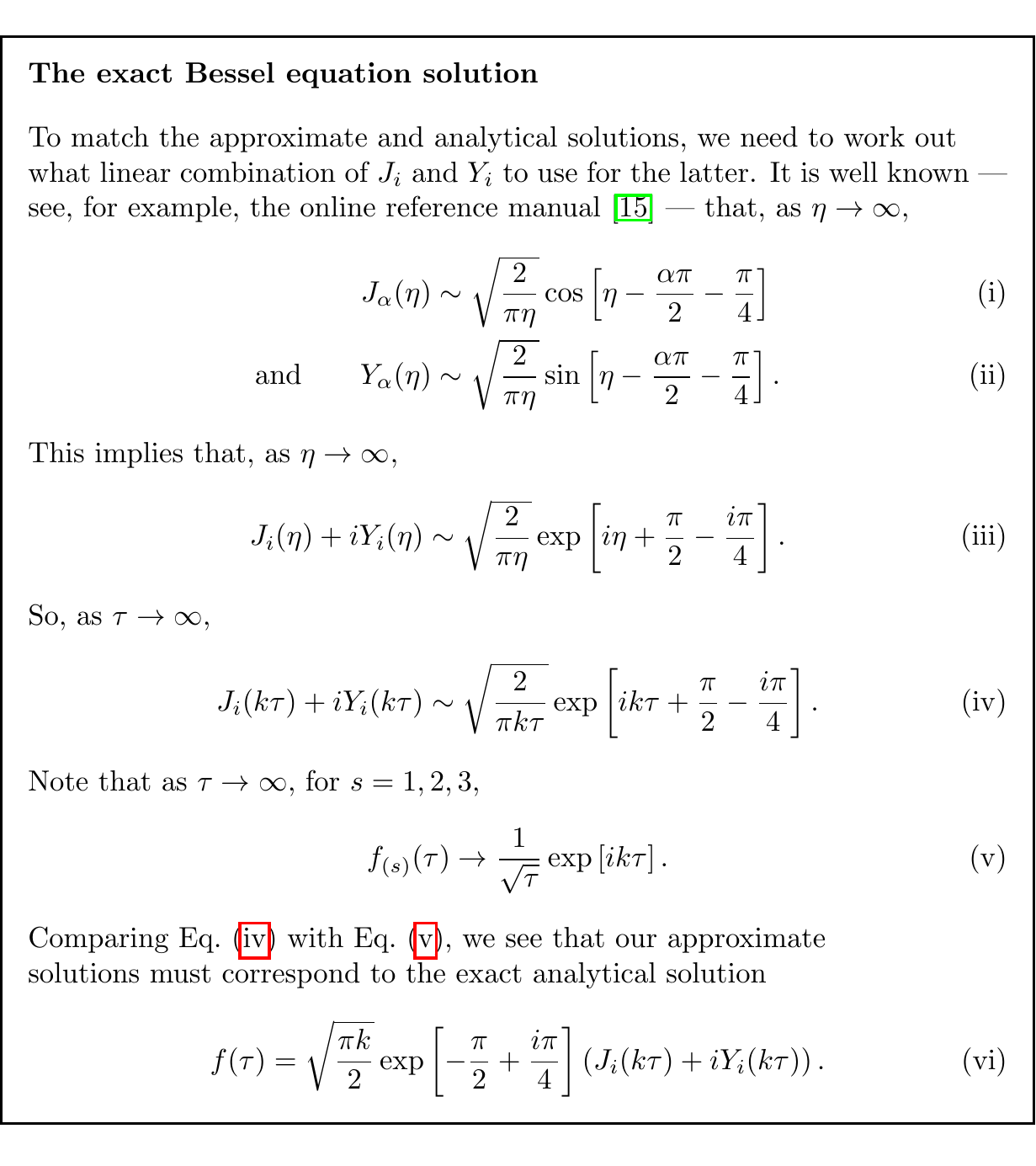}
\caption[Matching Bessel equation solutions]{Matching the approximate and exact solutions of our Bessel equation.}
\label{fig:matchingbessel}
\end{figure}

We increase $s$ by $1,$ to get $\underline{s=3}.$  Equating coefficients of $\kn^{2-3}=\kn^{-1}$ in \eref{eq:Besselapproxequation2}, gives us
\be 
\left(\omega_2'+2\omega_0\omega_3+2\omega_1\omega_2\right)+\frac{1}{\tn}\omega_2=0.
\ee
Using the values of $\omega_0,\omega_1$ and $\omega_2$ already calculated, this gives us
\be 
\left(-\frac{5i}{4\tn^3}+2i\omega_3-2\frac{1}{2\tn}\frac{5i}{8\tn^2}\right)+\frac{1}{\tn}\frac{5i}{8\tn^2}=0.
\ee
That results in
\be 
\omega_3=\frac{5}{8 \tn ^3},
\ee
which gives
\bas
f_{(3)}(\tn)&= \exp\left[\int \left(i\kn-\frac{1}{2\tn}+\frac{5i}{8\kn\tn^2}+\frac{5}{8\kn^2 \tn ^3}\right)d\tn\right]\\
&=\exp\left[i\kn\tn-\frac{1}{2}\log(\tn)-\frac{5i}{8\kn\tn}-\frac{5}{16\kn^2 \tn ^2}\right]\\
&= \frac{1}{\sqrt{\tn}}\exp\left[i\kn\tn-\frac{5i}{8\kn\tn}-\frac{5}{16\kn^2 \tn ^2}\right].
\eas
Figure~\ref{fig:FigBesselcompare}'s green line shows that this is yet a further improvement: $f_{(3)}$ is very close to the analytical solution for $\kn\tn\gtrsim 1.5.$

Whatever value of $\smax$ we choose, we expect the approximation to fail for some small enough $\tn$ because the approximation we have used involves a series in $\kn^{-1}$, and this fails once the value of $\kn^{-1}$ gets too large relative to $\tn$. $\kn\tn$ needs to be big enough to make the power series well behaved.

\section{The double power series approximation to third order}
\label{sec:double-power-series-for-perturbations}

In this appendix, we work through the double power series solution for Section~\ref{sec:an-approximate-solution-for-first-order-scalar-fast-modes-in-the-flat-radiation--matter-model} explicitly to third order, following the algorithm set out in Figure~\ref{fig:Doublepowerseries algorithm}.  Circled numbers refer to the steps of that algorithm.\\ \\
\vspace{6 pt}\noindent\textbf{\nc{1}}\nopagebreak \\ \nopagebreak
We take $\smax=3.$  In terms of the discussion at the end of Section~\ref{sec:introducing-the-double-power-series-method}, this gives a reasonable balance between simplicity~--- having relatively few terms~--- while, as we shall see in Figures~\ref{fig:Figrm100} and \ref{fig:Figrm100error}, being a valid approximation not too long after horizon--crossing and, once it is a valid approximation, having small errors.\\ \\
\vspace{6 pt}\noindent\textbf{\nc{2} to \nc{5}}\nopagebreak \\ \nopagebreak
Set $\underline{s=0}.$ We have
\be 
\matA^{-1}\matB[-2]=\begin{pmatrix}
	1&0\\
	4&1
\end{pmatrix}
\begin{pmatrix}
	0&0\\
	\frac{4}{3}&\frac{1}{3}
\end{pmatrix}
=\begin{pmatrix}
	0&0\\
	\frac{4}{3}&\frac{1}{3}
\end{pmatrix}.
\ee
It is easy to see that
\be
\vecp_0=\begin{pmatrix}
	0\\1
\end{pmatrix}
\ee is an eigenvector, with eigenvalue $\infrac{1}{3}$ and that the other eigenvalue is $0$.

The $0$ eigenvalue corresponds to the \Meszaros modes and the $\infrac{1}{3}$ eigenvalue to the oscillating modes.  Accordingly, as is intuitively plausible from the physics, the \Meszaros modes are particularly associated with pressureless matter and the oscillating modes with radiation pressure\footnote{To make this notion concrete and precise, suppose that, instead of confining ourselves to the case of radiation and pressureless matter, we had two fluids, $F_1$ and $F_2$, obeying respective equations of state, $P=w_\alpha\rho,$ with $w_\alpha$ constant for $\alpha=1,2.$ Then, we can do calculations similar to those which resulted in Eqs.~\eqref{eq:rgeq1c} and~\eqref{eq:rgeq2c} to get a pair of second order differential equations in the $F_1$ perturbation and $\psi$. It can then be shown that the eigenvalues of the resulting matrix $\matA^{-1}\matB[-2]$ would be $w_1$ and $w_2$. The $F_1$ perturbations would be the eigenvector of these equations associated with the eigenvalue $w_1$.  Note that there is no asymmetry between the two fluids here: we could, alternatively, derive a pair of second order differential equations in the $w_2$ fluid and $\psi,$ for which the $w_2$ perturbation would be the eigenvector with eigenvalue $w_2$. As usual, the Bianchi identities guarantee that there are only four independent modes.}.

Note that in Eqs.~\eqref{eq:rgeq1c} and~\eqref{eq:rgeq2c}, in accordance with step~\tnc{4}, we choose our eigenvalue as $\mu=\infrac{1}{3}.$  Following step~\tnc{5}, we set $\omega_0=\infrac{i}{\sqrt{3}}.$
\\ \\
\vspace{6 pt}\noindent\textbf{\nc{6}}\nopagebreak \\ \nopagebreak
The algorithm tells us to pick $\vecp_0$ to be an eigenvector of $\matA^{-1}\matB[-2]$ corresponding to the eigenvalue $\infrac{1}{3}.$ So we choose
\be
\vecp_0=\begin{pmatrix}
0\\
1
\end{pmatrix}.
\ee  

We can also look at this same calculation in a slightly different way, in terms of equating coefficients of $\kn^{2-0}=\kn^2$ in \eref{eq:doublepowerseriesapproxequation}. Most of the terms in \eref{eq:doublepowerseriesapproxequation} do not generate coefficients of $\kn^2$: only the $\matA\vecp\omega^2$ and the $\matB[-2]\, \kn^2\vecp$ terms contribute.  The resulting equation for $\vecp_0$ is
\bml
\left[\omega_0^2\matA+\matB[-2]\right]\vecp_0
\\
=\left[\omega_0^2\begin{pmatrix}
1&0\\
-4&1
\end{pmatrix}
+\begin{pmatrix}
0&0\\
\frac{4}{3}&\frac{1}{3}
\end{pmatrix}\right]\vecp_0=\vec{0}.
\eml
Solving the matrix equation, we again find we can set $\omega_0=\frac{i}{\sqrt{3}},$ and that
$\vecp_0$ can have no $\psi$ component; it only has a $\dr$ component, which we choose to be $1$.
\\ \\
\vspace{6 pt}\noindent\textbf{\nc{7}}\nopagebreak \\ \nopagebreak
This gives us our lowest order approximation
\be 
\vecf_{(0)}(\tn)=\begin{pmatrix}
0\\
1
\end{pmatrix}
\exp\left[\frac{i\kn\tn}{\sqrt{3}}\right].
\ee
In this approximation, $\psi=0,$ but  $\dr\ne 0,$ so \eref{eq:rgeq1c} suggests that this does not approximate the exact solution very well.\\ \\
\vspace{6 pt}\noindent\textbf{\nc{8} to \nc{11}}\nopagebreak \\ \nopagebreak
These steps are analogous to those in Appendix~\ref{sec:Bessel-calculation} for the Bessel example, but involve vector equations. We increase $s$ by $1$ to get $\underline{s=1}.$  We equate coefficients of $\kn^{2-1}=\kn$ in \eref{eq:doublepowerseriesapproxequation}. We use square brackets to group together terms which come from a single term of \eref{eq:doublepowerseriesapproxequation}, and have
\be \label{eq:radns=1}
\matA\left[\vec{p}_0 2\omega_0\omega_1 + \vec{p}_1 \omega_0^2 \right]+\matC\vec{p}_0\omega_0 +\matB[-2]\,\vec{p}_1=\vec{0}.
\ee
Note the rule that for each term
\bml \label{eq:equaterule}
\text{(Total of subscripts in the term)}
\\
-\text{(Total power of $\omega$s in the term)}=s-2.
\eml
The subscripts totalled include those for the $\vecp,\  \omega$ and also the $\matB$ terms~--- and is the underlying motivation for our choice of subscripts $-2$ and $0$ for the $\matB$s.  The rule follows from looking at coefficients of $\kn^{-1}$ from the double power series expression of \eref{eq:vxexpn}.

It is significant that $\vec{p}_1$ occurs only in the combination of terms $\left[\matA\omega_0^2+\matB[-2]\right]\vec{p}_1$.  Indeed, at every iteration $s,$ we get $\vec{p}_s$ making its first appearance in the algorithm in the combination of terms $\left[\matA\omega_0^2+\matB[-2]\right]\vec{p}_s.$   This gives an undetermined degree of freedom for each $\vec{p}_s$ which is parallel to the original eigenvector $\vec{p}_0,$ which satisfied the equation $\left[\matA\omega_0^2+\matB[-2]\right]\vec{p}_0=0.$ Step~\tnc{4} of our algorithm includes the assumption that the eigenvalue $-\omega_0^2$ is not a repeated root of the characteristic equation, and so the eigenspace is of dimension one: in other words, there is a single undetermined degree of freedom for each $\vec{p}_j.$ In accordance with step~\tnc{9} in the algorithm, we set this undetermined degree of freedom to zero. The rationale for this is that the choice of this degree of freedom will not affect the approximation, so it is best to choose the degree of freedom to keep the expression for the approximation as simple as possible.  

In what follows, we will write, for any $s,$
\be \label{eq:ps}
\vec{p}_s=
\begin{pmatrix}
	p_s^{(\psi)}\\[6pt]
	p_s^{(\dr)}
\end{pmatrix}.
\ee
Putting $\matA$, $\matB[-2]$, $\vec{p}_0$ and $\omega_0$ into \eref{eq:radns=1}  we find
\be
\begin{pmatrix}
-\frac{1}{3}p_1^{(\psi)}\\[6pt]
-\frac{8}{3}p_1^{(\psi)}+\frac{2}{\sqrt{3}}i\omega_1
\end{pmatrix}
=\vec{0}.
\ee
Hence
\be
p_1^{(\psi)}=0 \qquad\text{and}\qquad\omega_1=0.
\ee\\ 
\indent We find, however, that $p_1^{(\dr)}$ is undetermined, so,  along lines discussed following \eref{eq:Bessel-f3main}, and in accordance with step~\tnc{9}, we set $p_1^{(\dr)}=0$. This gets us
\be
\vec{p}_1=\vec{0}.
\ee
Because for $s\ge 1,\ \vec{p}_s$ always makes its first appearance in the algorithm in the combination of terms $\left[\matA\omega_0^2+\matB[-2]\right]\vec{p}_s,\ p_s^{(\dr)}$ is always undetermined and therefore set to $0.$

Because $\omega_1=0$ and $\vecp_1=\vec{0},$ our approximation is unchanged by this $s=1$ iteration of the algorithm. We have
\be 
\vecf_{(1)}(\tn)=\begin{pmatrix}
	0\\
	1
\end{pmatrix}
\exp\left[\frac{i\kn\tn}{\sqrt{3}}\right]=\vecf_{(0)}(\tn).
\ee \\ 
\indent In accordance with step~\tnc{8}, we next increase $s$ by $1,$ to get $\underline{s=2}.$ We equate coefficients of $\kn^{2-2}=\kn^0=1$ in \eref{eq:doublepowerseriesapproxequation}. This gives the equation
\bml
\matA\left[\vec{p}_2\,\omega_0^2+\vec{p}_0\left(2\omega_0\omega_2+\omega_1^2\right)\right]
\\
+\matC\left[\vec{p}_0\omega_1+\vecp_1\omega_0\right]+\left(\matB[0]\,\vec{p}_0+\matB[-2]\,\vec{p}_2\right)=0.
\eml
Using the values of $\vec{p}_0$, $\omega_0$, $\vec{p}_1$ and $\omega_1$ found above, this gives us
\be
\omega_2=-\frac{32i\sqrt{3}}{\tn^2\left(4+\tn\right)^2}
\ee
and
\be\vec{p}_2=
\begin{pmatrix}
-\frac{24}{\tn^2\left(4+\tn\right)^2}\\[6pt]
0
\end{pmatrix},
\ee
where, as noted above, $p_2^{(\dr)}$ was found to be undetermined, leading us to set it to $0$.

To get $\vecf_{(2)},$ we now have to do the integration
\bml
\int \kn^{-1}\omega_2\,d\tn 
=\kn^{-1}\int{-\frac{32i\sqrt{3}}{\tn^2\left(4+\tn\right)^2}}\,d\tn
\\
=-i\kn^{-1}\sqrt{3}\int\left\lbrace\frac{2}{\tn^2}+\frac{2}{\left(4+\tn\right)^2}-\frac{1}{\tn}+\frac{1}{4+\tn} \right\rbrace\,d\tn
\\
=-i\kn^{-1}\sqrt{3}\left\lbrace -\frac{2}{\tn}-\frac{2}{4+\tn}-\log(\tn)+\log(4+\tn)\right\rbrace
\\
=\frac{4 i \sqrt{3} (2+\tn)}{\kn \tn  (4+\tn)}+\frac{i \sqrt{3}}{\kn}\log\left[\frac{\tn}{(4+\tn)}\right],
\eml
where in the second line we have adopted the standard approach to such integrations of splitting our original fraction into a series of partial fractions. 

We now have
\bml 
\vecf_{(2)}(\tn)
\\=
\begin{pmatrix}
	-\frac{24}{\tn^2\left(4+\tn\right)^2}\\[6pt]
	1
\end{pmatrix}
\left(\frac{\tn }{4+\tn}\right)^{\frac{i \sqrt{3}}{\kn}}\exp\left[\frac{i\kn\tn}{\sqrt{3}}+\frac{4 i \sqrt{3} (2+\tn)}{\kn \tn  (4+\tn)}\right]
.\\ 
\eml 

We increase $s$ by $1$ again, to get $\underline{s=3}.$  We equate coefficients of $\kn^{2-3}=\kn^{-1}$  in \eref{eq:doublepowerseriesapproxequation}.  The resulting equation has more terms than for $s=1,$ because we now pick up derivative terms from \eref{eq:doublepowerseriesapproxequation}: we have
\bml \label{eq:f3}
\matA\left(2\vecp_2'\omega_
0+\vecp_0\omega_2'+\left[\vec{p}_3\,\omega_0^2+\vec{p}_22\omega_0\omega_1
\right.\right.\\ \left.\left.
+\vecp_1\left(\omega_1^2+2\omega_0\omega_2\right)+\vecp_0\left(2\omega_0\omega_3+2\omega_1\omega_2\right)\right]\right)\\+\matC\left[\vec{p}_0\omega_2+\vecp_1\omega_1+\vecp_2\omega_0\right]
\\
+\left(\matB[0]\,\vec{p}_1 +\matB[-2]\,\vec{p}_3\right)=0.
\eml
The rule from \eref{eq:equaterule} is very helpful in constructing the above expression.  As before, square brackets group together terms which come from a single term of \eref{eq:doublepowerseriesapproxequation}.

Substituting the results already obtained for $\vecp_0,\,\omega_0,\,\vecp_1,\,\omega_1,\,\vecp_2$ and $\omega_2$ into \eref{eq:f3}, we get~--- after some algebra~---
\begin{align}
\omega_3&=0\\
\text{and}\qquad\vec{p}_3&=
\begin{pmatrix}
\frac{48\sqrt{3}i\left(2+\tn\right)}{\tn^3\left(4+\tn\right)^3}\\[6pt]
0
\end{pmatrix}.
\end{align}

This gives us the approximate solution $\vecf_{(3)},$ which is
\begin{align} \label{eq:approxrad}
\begin{split}
\dr(\tn)&=\left(\frac{\tn }{4+\tn}\right)^{\frac{i \sqrt{3}}{\kn}}\exp\left[ \frac{i \kn \tn }{\sqrt{3}}+\frac{4 i \sqrt{3} (2+\tn)}{\kn \tn  (4+\tn)} \right],\\ \\
\psi(\tn)&=-\frac{24}{\kn^2 \tn ^2 (4+\tn)^2}\left\lbrace1-\frac{2 i\sqrt{3} \left( 2+\tn \right)}{\kn \tn (4+\tn)}\right\rbrace\dr(\tn).
\end{split}
\end{align}
Recall from \eref{eq:tauunit} that, in the above, $\tn$ is conformal time measured in units of $\tunit,$ and $\kn$ is the co--moving wave--number measured in units of $\infrac{\Hrm}{\sqrt{2}}$.

\begin{widetext}
\section{The double power series approximation to seventh order}\label{sec:app-seventh}
The approximate solution set out in this appendix is derived using the same approach as in  Section~\ref{sec:an-approximate-solution-for-first-order-scalar-fast-modes-in-the-flat-radiation--matter-model} but going to seventh order, using our \emph{Mathematica} notebook available at \rfcite{Github}.  In particular,
$\dm$ is obtained by using the ninth order $\dr$ and $\psi$ approximations in \eref{eq:matter-tau} and dropping any powers of $\kn^{-n}$ with $n>7$ from the factor in braces in \eref{eq:dm7}.

\begin{multline}\label{eq:7-r}
\dr(\tn)=
\exp \Bigg[\frac{i \kn \tn }{\sqrt{3}}
+{\frac{i \sqrt{3} \left(16 \kn^4+78 \kn^2+819\right)}{16 \kn^5}} \log\left[\frac{\tn }{\tn +4}\right]
+\frac{4 i \sqrt{3} (\tn +2)}{\kn \tn  (\tn +4)}
\\
+\frac{i \sqrt{3} \left(39 \tn ^5+390 \tn ^4+1144 \tn ^3+624 \tn ^2-960 \tn -256\right)}{2 \kn^3 \tn ^3 (\tn +4)^3}
-\frac{288 \left(11 \tn ^2+44 \tn +16\right)}{\kn^4 \tn ^4 (\tn +4)^4}\\
+\frac{3 i \sqrt{3} \left(1365 \tn ^9
+24570 \tn ^8+171080 \tn ^7+560560 \tn ^6+797888 \tn ^5+232960 \tn ^4-640 \tn ^3+928000 \tn ^2+814080 \tn +147456\right)}{20 \kn^5 \tn ^5 (\tn +4)^5}\\
+\frac{432 \left(987 \tn ^4+7896 \tn ^3+18320 \tn ^2+10112 \tn +512\right)}{\kn^6 \tn ^6 (\tn +4)^6}
\Bigg], 
\end{multline}

\begin{multline}\label{eq:7-psi}
\psi(\tn)=-\frac{24}{\kn^2 \tn ^2 (\tn +4)^2}
\Bigg\lbrace 1-\frac{2 i \sqrt{3} ( \tn+2)}{\kn \tn ( \tn+4)}
-\frac{36}{\kn^2 \tn (\tn+4)}
+\frac{216 i \sqrt{3} ( \tn+2)}{\kn^3 \tn^2 (\tn+4)^2}\\
+\frac{72 \left(63 \tn ^4+504 \tn ^3+1192 \tn ^2+736 \tn +64\right)}{\kn^4 \tn ^4 (\tn +4)^4}\\
-\frac{576 i \sqrt{3} \left(63 \tn ^5+630 \tn ^4+2158 \tn ^3+2868 \tn ^2+1168 \tn +64\right)}{\kn^5 \tn ^5 (\tn +4)^5}\Bigg\rbrace \,\dr(\tn),
\end{multline}

\begin{multline} \label{eq:dm7}
\dm(\tn)=-\frac{144}{\kn^2 \tn ^2 (\tn +4)^2}
\Bigg\lbrace
1-\frac{5 i \sqrt{3} (\tn +2)}{\kn \tn  (\tn +4)}
-\frac{6 \left(19 \tn ^2+76 \tn +48\right)}{\kn^2 \tn ^2 (\tn +4)^2}
\\
+\frac{72 i \sqrt{3} \left(13 \tn ^3+78 \tn ^2+132 \tn +56\right)}{\kn^3 \tn ^3 (\tn +4)^3}
-\frac{24 \left(63 \tn ^4+504 \tn ^3+1164 \tn ^2+624 \tn +128\right) (\tn +2)^2}{\kn^4 \tn ^5 (\tn +4)^5}
\\
-\frac{144 i \sqrt{3} \left(294 \tn ^5+2940 \tn ^4+11183 \tn ^3+20058 \tn ^2+16776 \tn +5152\right)}{\kn^5 \tn ^5 (\tn +4)^5}
\Bigg\rbrace
\,\dr(\tn).
\end{multline}
As for \eref{eq:approx}, here $\tn$ is the value of the conformal time, when measured in units of $\tunit,$ and $\kn$ is the value of the co--moving wave--number, when measured in units of $\kunit.$

The above solutions are plotted for $\kn=10$ in Figure~\ref{fig:Figrm10}, with the errors shown in Figure~\ref{fig:Figrm10error}. Lower wave--numbers tend to be more testing to approximate, so we chose the value $\kn=10$ to demonstrate the accuracy of the seventh order approximation.

\begin{figure}
\centering
\includegraphics[width=1\linewidth]{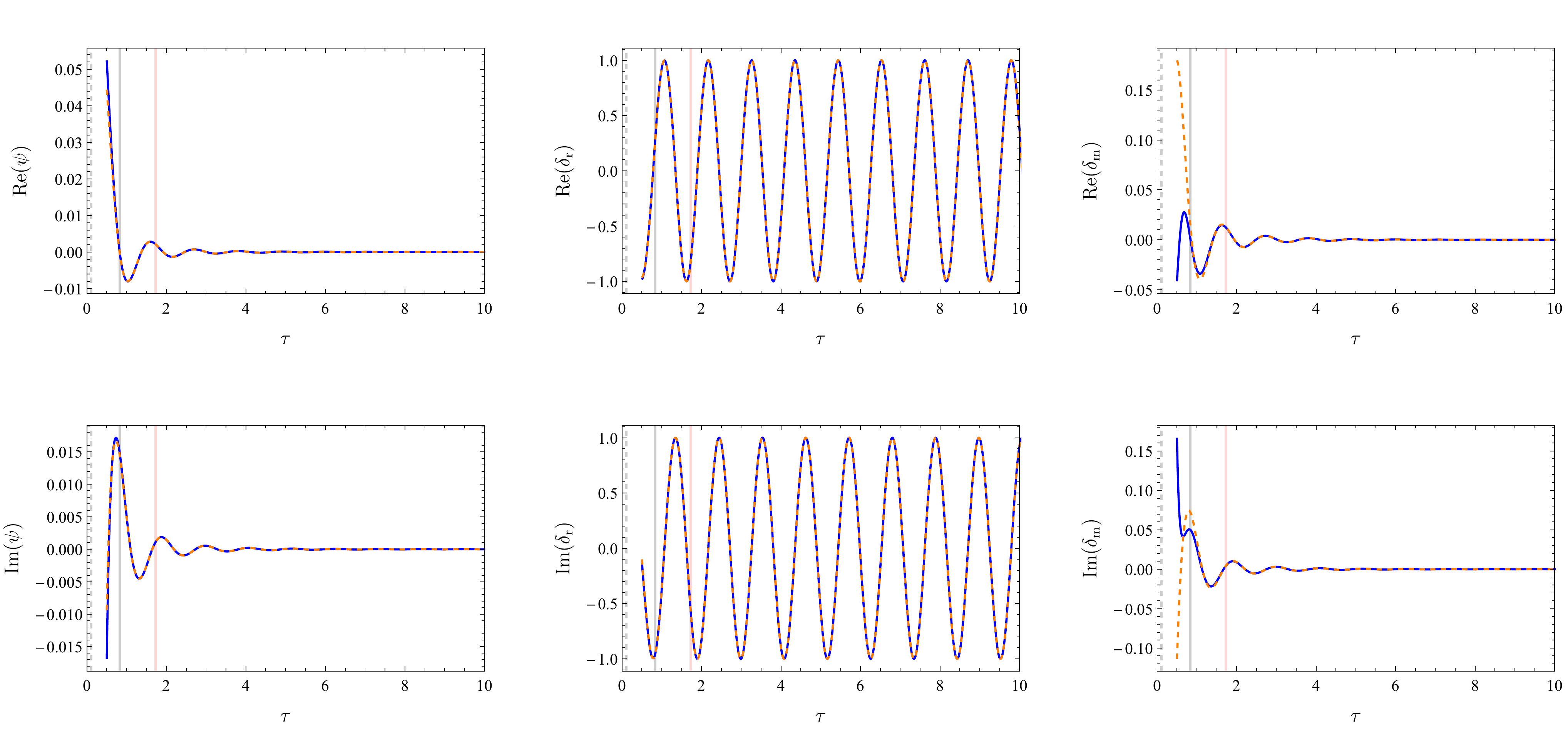}
\caption[Approximate solution for a flat radiation--matter universe, $\kn=10$]{The solid blue line is the seventh order double power series solution with $\kn=10.$ The orange dashed line is the numerical solution. The pink vertical line shows the time of photon--baryon decoupling. Other details are as in Figure~\ref{fig:Figrm100}.} 
\label{fig:Figrm10}

\centering
\includegraphics[width=1\linewidth]{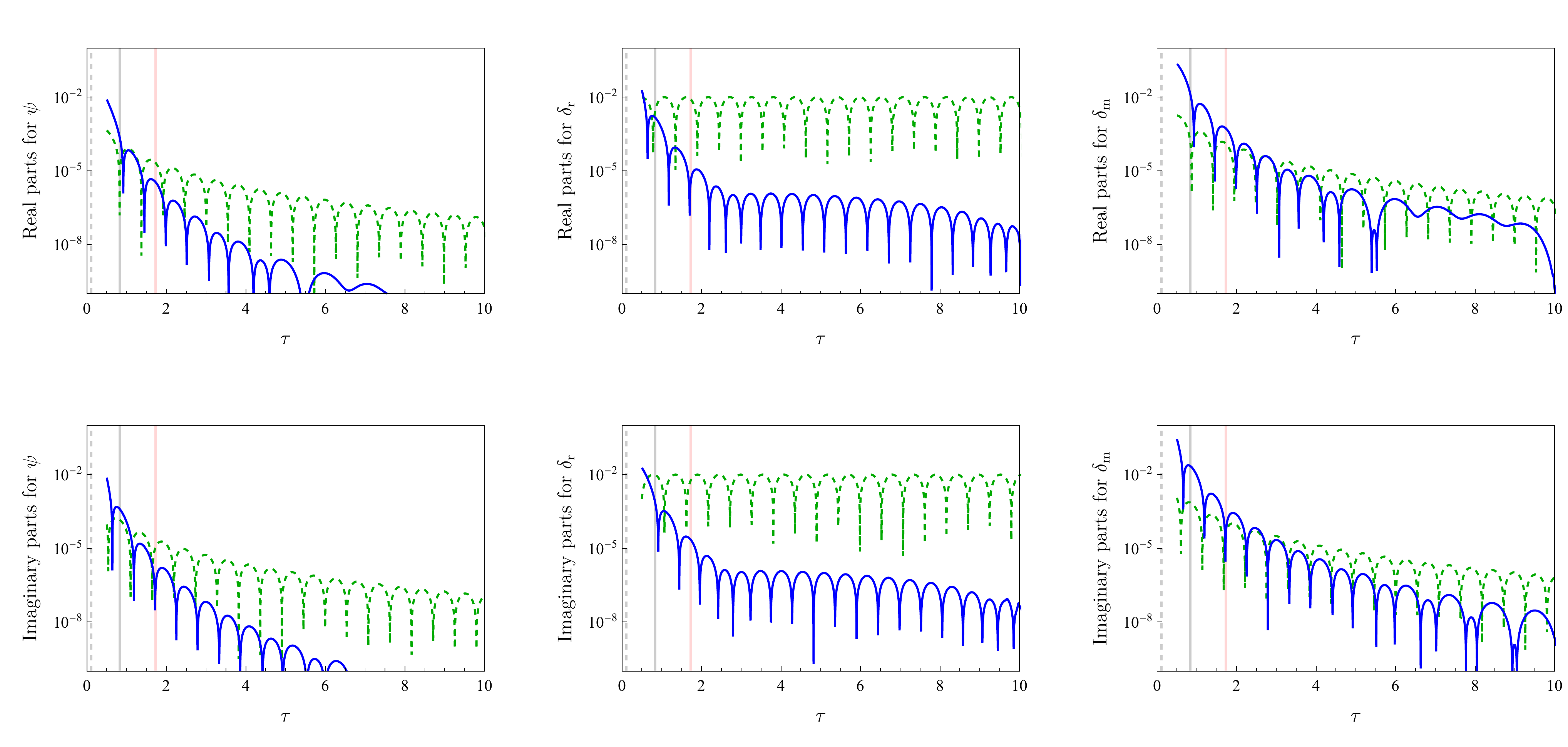}
\caption[Error of the approximate solution for a flat radiation--matter universe, $\kn=10$]{This figure shows the errors in the approximations for $\kn=10$ of Figure~\ref{fig:Figrm10}. In brief, the solid blue line is the relative error of the seventh order double power series solution for $\kn=10,$ compared with the numerical solution.  The green dashed line is $1\%$ of that numerical solution.  See the caption to Figure~\ref{fig:Figrm100error} for more details.} 
\label{fig:Figrm10error}
\end{figure}
\end{widetext}

\end{document}